\documentclass[aps,prb,groupedaddress,notitlepage]{revtex4-1}

\usepackage{graphicx}
\usepackage{amsmath}
\usepackage{braket}
\usepackage{appendix}
\usepackage{bbm}
\usepackage[T1]{fontenc}
\begin{document}


\title{Efficient vertex parametrization for the constrained functional renormalization group for effective low-energy interactions in multiband systems}


\author{Carsten Honerkamp}
\email[]{honerkamp@physik.rwth-aachen.de}
\affiliation{
Institute for Theoretical Solid State Physics, RWTH Aachen University,
D-52056 Aachen and JARA - Fundamentals of Future Information
Technology\\
}

\date{May 4, 2018}

\begin{abstract}
We describe an efficient approximation for the electron-electron interaction in the determination of the low-energy effective interaction in multiband lattice systems. By using ideas for channel decomposition, form-factor expansion and the truncated-unity technique we describe the interaction as arising from the non-local and orbital-dependent coupling of  particle-hole and particle-particle bilinears formed by fields residing in the same one or two orbitals. This allows us to employ the constrained functional renormalization group (cfRG) with a suitable momentum and frequency discretization. The approach gives insights into the non-local screening of spin and charge interactions when bands away from the Fermi level are integrated out. Specifically, we compute the effective low-energy interactions in the low-energy target band of a three-band model with onsite and non-local bare interactions. We show that the cfRG adds important features to the effective target-band interaction that cannot be found using the constrained random phase approximation (cRPA).
\end{abstract}


\maketitle

\setlength{\parindent}{0pt}
\section{Introduction}
The constrained random phase approximation\cite{aryasetiawan,miyake,imada,sasioglu} (cRPA) is a very useful scheme for determining the effective interaction in low-energy models for electrons in solids. Starting with a band structure on a wider energy scale, it allows one to take into account efficiently the screening of the interactions that act in the low-energy window by the electrons in the bands outside this window. 
This screening is an important physical effect. Accounting for it paves the way for parameter-free calculations\cite{sasioglu} of properties of correlated materials. As expressed by its naming, the cRPA is an approximate scheme, perturbative in the electron-electron interactions, that consists in selecting a certain class of diagrams. It amounts to an infinite-order summation of an appropriately defined polarization function. The constraint consists in disallowing contributions to this polarization that are purely due to the low-energy bands (also called target bands). This means that the allowed contributions are those within the high-energy bands only and those between the high-energy and low-energy bands. 

As with any approximation, it is valuable to know how good it is and whether there are relevant corrections terms to it. This holds in particular for the cRPA in its usual realm, where no small parameter like large $N$ or small $q/k_F$ exists that would render some type of control over this approximation. In  two previous works\cite{honerkampcfRG,kinza}, we have proposed to use a constrained functional renormalization group scheme (that we from now on call cfRG)  to extend the cRPA and hence to include additional diagrams that are neglected in the cRPA into the calculation of the effective interactions. The cfRG is an adaptation of the general functional renormalization group framework for interacting fermions\cite{metzner} to the problem of tailoring effective target-band actions by integrating over the high-energy bands in multiband electron systems. We showed that at least in simple models\cite{kinza}, sizable corrections to cRPA can exist (but do not have to - this depends on the model). In order to interpret these corrections we analyzed the frequency and momentum dependence of these terms. However, these dependencies become quite rich, as in the cfRG the effective interaction of a translationally symmetric model depends on three momenta and three frequencies, in addition to possibly four band or orbital indices. Therefore the cfRG-treatment of more realistic models that embody more definite material properties is facing a bottleneck of how this wealth of information can be processed and evaluated efficiently. In this aspect, the cRPA is simpler, as its effective interaction only depends on one frequency and one momentum. In addition its results can be interpreted in terms of effective dielectric functions.

Regarding the complexity of the momentum and frequency dependence, recent fRG literature contains considerable progress. One important tool has become the channel decomposition\cite{husemann,husemann12,giering,wang} of the flowing interaction into pairing, charge ans spin channels (or linear combinations of the latter two). This decomposition can be motivated both by the structure of second order perturbation theory in the interactions as well as by the effective interactions found in the fRG for usual models. It allows to read the interaction as mediated by collective boson-like propagators that couple to pairing, charge or spin bilinears, depending on the respective channel.
In addition to that, a form-factor expansion\cite{husemann,husemann12,giering,wang} of the internal spatial structure of these coupling bilinears was found to converge quickly for standard settings\cite{lichtenstein,sanchez1,sanchez2}. As a benefit of these reformulations of the momentum structure, the fRG codes can be pushed to much higher momentum resolution and perform well on highly parallel computing infrastructures.
Regarding the frequency structures, the situation is slightly more complex because at least at stronger coupling\cite{rohringer12,kinza13} and in particular contexts\cite{vilardi}, a simple channel decomposition, as pioneered in zero-dimensional models\cite{karrasch} and recently also used in one dimension\cite{cla,markhof} can be problematic. Nevertheless, for weaker coupling, the channel decomposition with some additional tweaking was found to make sense in standard setups empirically\cite{reckling}. It was also shown that a so-called static channel-coupling approximation for a channel-decomposed one-frequency parametrization of the frequency depdendence is quantitatively reasonable as long as the initial interaction is not retarded and no strong retardations are generated during the flow. Note also that form-factor expansions for the frequency dependence are being worked out currently\cite{yirga}. These could make the one-frequency parametrizations more robust without increasing the numerical effort too strongly. Furthermore, efficient treatments of the high-frequency asymptotics of the vertex have been considered\cite{wentzell}.

In this work we present a channel-decomposed fRG scheme for the computation of the effective interactions in a low-energy target band when bands away from the Fermi level are integrated out. Besides restricting the form-factor expansion to local bilinears, we also choose to concentrate on the orbital-diagonal bilinears. These bilinears are the ones that interact in the bare action and hence neglecting orbital non-diagonal bilinears appears to be justifiable unless particular evidence for the generation of or interest in new terms is present. Of course the formalism can be expended to include those bilinears as well as non-local form factors. Our goal here is to showcase a numerically fast core scheme with diagonal local bilinears that, despite the simplifications, produces more and interestingly different renormalizations compared to the cRPA.
 
\section{Model}
We assume that our model on the wider energy scale of a few tens of eV is spin-rotational symmetric. This assumption can also be relaxed (see Ref. \onlinecite{schober} for a description of channel-decomposed fRG without spin-rotational invariance), but for the sake of this paper we keep spin rotational invariance alive as this simplifies the formulae. We further assume that the model can be formulated in a Wannier basis. The Wannier orbitals within a unit cell shall be indexed by $o$ such that we have a Hamiltonian of the following type,
\begin{eqnarray}
H &=& \sum_{\vec{k},s,s' \atop o,o'}h_{oo'} (\vec{k} )c_{\vec{k},s,o}^\dagger c_{\vec{k},s,o'} \nonumber \\&&+ \frac{1}{2N}  \sum_{\vec{k}_1,\vec{k}_2,\vec{k}_3,s,s' \atop o_1,o_2,o_3,o_4} V_{o_1o_2o_3o_4} (\vec{k}_1,\vec{k}_2,\vec{k}_3) 
c_{\vec{k}_3,s,o_3}^\dagger c_{\vec{k}_4,s',o_4}^\dagger c_{\vec{k}_2,s',o_2} c_{\vec{k}_1,s,o_1} \, . \label{hamilton} \\
\end{eqnarray}  
The wavevectors $\vec{k}$ live in the first Brillouin zone of the lattice with $N$ unit cells and periodic boundary conditions.
$\vec{k}_4$ is determined by $\vec{k}_1+\vec{k}_2-\vec{k}_3$ modulo  reciprocal lattice vectors. The kinetic matrix $h_{oo'} (\vec{k})$ contains the hopping amplitudes which are obtained from the band structure or from the overlap matrix elements between the Wannier functions.  
\subsection{Band structure of three-orbital test model}
\begin{figure}
 \includegraphics[width=.48\columnwidth]{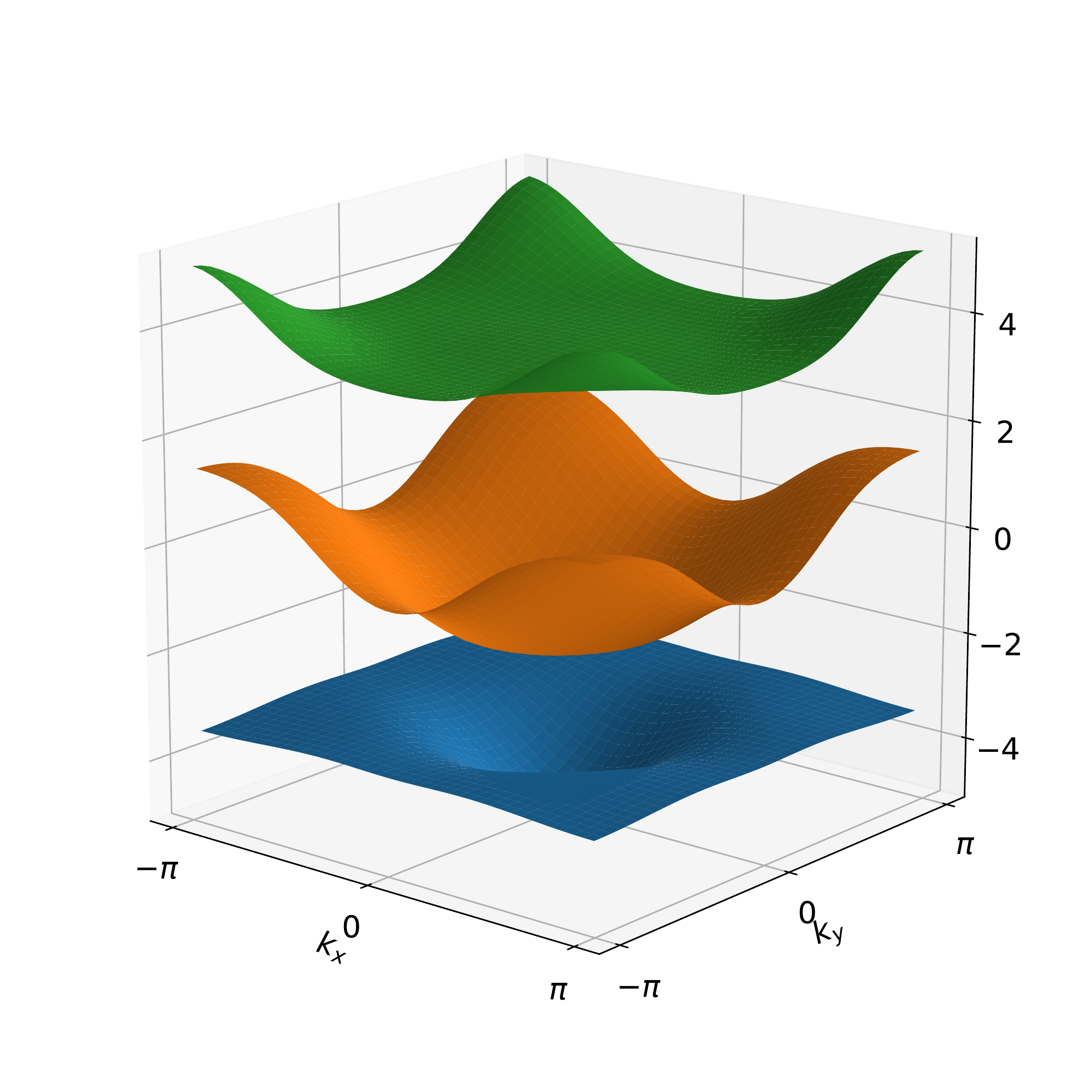}%
 \caption{Band structure of the three-orbital model described in the text in Eq. \ref{3band} with parameters $t_{11}= 1$, $t_{12}=2$, $t_{13}=0.5$, $t_{22}=t_{33}=-0.1$, $\epsilon_{1} = 0$, $\epsilon_2=-\epsilon_3=2$. The middle band crosses the Fermi level and is called target band. The band filling of the central band is roughly $40\%$. }
  \label{modelfig}
\end{figure}

For the testing of the formalism developed below we choose a three-orbital model with a kinetic matrix
\begin{equation}
\label{3band}
\hat h = \begin{pmatrix}
 \epsilon_1- 2 t_{11}  (\cos k_x+ \cos k_y) & t_{12} & t_{13} \\
     t_{12}  & \epsilon_2- 2 t_{22}  (\cos k_x+ \cos k_y)  & t_{23}   \\
t_{13} & t_{23}      & \epsilon_3 - 2 t_{33}  (\cos k_x+ \cos k_y)   
\end{pmatrix} \, .
\end{equation}
Typical parameter choices are (in some sensible energy unit) $t_{11}= 1$, $t_{12}=2$, $t_{13}=0.5$, $t_{22}=t_{33}=-0.1$, $\epsilon_{1} = 0$, $\epsilon_2=-\epsilon_3=2$. This example gives rise to the bands shown in Fig. \ref{modelfig}.
\subsection{Multi-orbital interactions} \label{multiorbV}
In general the Coulomb interaction expressed in the Wannier basis can be quite complex, but in many cases it is reasonable to focus on to a two-center approximation in the orbital indices, i.e. consider nonzero terms where only two physical orbitals are involved. For these terms we consider the following standard interactions:
\begin{itemize}
  \item Intraorbital density-density interactions of the type
\begin{equation}
V_{o_1o_2o_3o_4} (\vec{k}_1,\vec{k}_2,\vec{k}_3)  = \delta_{o_1o_3} \delta_{o_2o_4}  V^{\rho}_{o_1o_2}  (\vec{k}_1-\vec{k}_3) \, . 
\label{vcoulomb}
\end{equation}
Terms like these arise via Fourier transformation from direct matrix elements of the Coulomb interaction with Wannier states $w_{o}(\vec{r}-\vec{R})$, centered at positions $\vec{r}_{o}$ in the unit cell indexed by Bravais lattice vectors $\vec{R}$,
\begin{equation}
V^{\rho}_{o_1o_2} (\vec{R} ) = V_c \int d^3r \int d^3r' \, \frac{|w_{o_1} (\vec{r} + \vec{r}_{o_1})|^2|w_{o_2} (\vec{r}' + \vec{r}_{o_2} -\vec{R})|^2 }{|\vec{r} + \vec{r}_{o_1} - \vec{r}' -\vec{r}_{o_2} +\vec{R}| }
\end{equation}
Included in this are for $\vec{R}=0$ the onsite intraorbital (for $o_1=o_2$) and interorbital   (for $o_1\not= o_2$) interactions. The non-local terms for $\vec{R} \not =0$ will still have some orbital dependence for small $\vec{R}$ if the orbitals are not placed at the same center position in the unit cell or have strongly varying spatial extent. However, the Coulomb tail for $\vec{R} $ much larger than the lattice constant and the $\vec{r}_o$,  will to good approximation become independent of the orbital index, and only depend on the distance $|\vec{R}|$. Hence, the small-$\vec{q}$ part $V^{\rho}_{o_1o_2}  (\vec{q}) $ will also be nearly orbital-independent. In the model calculations below we will include the Coulomb tail by the Hamiltonian
\begin{equation} \label{hlr}
H_{\mathrm{Coul}} =\frac{V_c}{2} \sum_{\substack{\vec{R}\not=\vec{R}' \\ o_1,o_2,\sigma,\sigma'}}  \frac{ n_{\vec{R},o_1,\sigma} n_{\vec{R}',o_2,\sigma'}}{|\vec{R}-\vec{R}'|} \, e^{-|\vec{R}-\vec{R}'|/\lambda} \, , 
\end{equation}
assuming $ \vec{r}_{o_1} = \vec{r}_{o_2} $ inside the unit cell and $\vec{R},\vec{R}'$ running over the unit cell positions. We have added in by hand an exponential decay factor with a screening length $\lambda$. This factor regularizes the divergence of the Fourier transform of (\ref{hlr}) and is hence needed in the numerical implementation. We usually take $\lambda=6$ lattice constants.
  \item Local intraorbital local density-density and spin-spin interactions are often presented in the form of the so-called Kanamori Hamiltonian (see, e.g., Refs.  \onlinecite{georges,piefke}), which reads (suppressing the site index for this onsite interaction)
  \begin{eqnarray} 
{H}_U &=& U \sum_{o_1} n_{o_1,\uparrow} n_{o_1,\downarrow} +
\sum_{\substack{o_1>o_2\\\sigma,\sigma'}} 
(U'-\delta_{\sigma \sigma'}J) n_{o_1,\sigma} n_{o_2,\sigma'}  \nonumber
\\
&& -\sum_{o_1 \neq o_2}
J\left (c^\dagger_{o_1,\downarrow}c^\dagger_{o_2,\uparrow}c^{\phantom{\dagger}}_{o_2,\downarrow}c^{\phantom{\dagger}}_{o_1,\uparrow}
+
c^\dagger_{o_2,\uparrow}c^\dagger_{o_2,\downarrow}c^{\phantom{\dagger}}_{o_1,\uparrow}c^{\phantom{\dagger}}_{o_2,\downarrow}
\right) \nonumber \\ 
&=& 
U \sum_{o_1} n_{o_1,\uparrow} n_{o_1,\downarrow} +
\sum_{\substack{o_1 \not= o_2\\\sigma,\sigma'}} 
\frac{U'}{2} n_{o_1,\sigma} n_{o_2,\sigma'} \nonumber
\\
&& + \frac{J}{2} \sum_{o_1 \neq o_2 \atop \sigma, \sigma' }
\left( c^\dagger_{o_1,\sigma}c^\dagger_{o_2,\sigma'}
c^{\phantom{\dagger}}_{o_1,\sigma'}c^{\phantom{\dagger}}_{o_2,\sigma}
+
c^\dagger_{o_2,\sigma} c^\dagger_{o_2,\sigma'}
c^{\phantom{\dagger}}_{o_2,\sigma'}c^{\phantom{\dagger}}_{o_1,\sigma} \right) 
 \,. \label{vkanamori}
\end{eqnarray} 
Here, $U$ and $U'$ are intra- and inter-orbital density-density interaction parameters and $J$ is the Hund's rule interaction. In cases with rotational invariance one has $U-U'=2J$. Note that also (\ref{vkanamori}) only involves two orbital indices $o_1$ and $o_2$, i.e. is an interaction of equal-orbital fermion bilinears. More general local density-density interactions $U_{oo'}$ or spin-spin interactions $J_{oo'}$ are also conceivable and do not lead to extra efforts in the approximations discussed below.
  
\end{itemize}
In general, even if the bare Hamiltonian of the model with the initial wide bandwidth does only contain these two-center interaction terms, the renormalization group flow that integrates out the high energy bands will generate all sorts of more complicated terms that depend on four orbital indices and that also exhibit dependencies on three momenta and frequencies. 
This complexity is hard to deal with for a true multiband situation with many bands. Below we propose approximation strategies to reduce the effort. 

\section{cRPA}
The cRPA\cite{aryasetiawan,miyake,imada,sasioglu} is usually formulated in the band picture and bands $b$ that do not belong to the target bands near the Fermi level are integrated out. Starting with the bare Coulomb repulsion, one computes an effective interaction $W^{\mathrm{cRPA}}(q)$ that acts between electron bilinears $\bar{c}_{b_3,k+q,s } c_{b_1,k,s}$ and $\bar{c}_{b_4,k'-q,s } c_{b_2,k',s}$, here expressed with fermion Grassmann fields that depend on frequency-momentum indices $k$,  target band indices $b_i$ in the low-energy space and spin projections $s$ and $s'$. The corresponding Bethe-Salpeter equation is shown in Fig. \ref{5diags} a).

For the context of this paper, the orbital representation is more appropriate. Then we can group the orbital indices $o_1$ and $o_3$ belonging to external legs with spin $s$ (say on the left side of the rectangular vertex representing the bare Coulomb interaction $V_{o_1o_2o_3o_4} (\vec q) $ in Fig. \ref{5diags} a) or Fig. \ref{VPCD} a)) to a double index and those of the other two legs with spin $s'$, $o_2$ and $o_4$, to another double index.
The cRPA effective interaction is, using hats $\hat{\ }$ to indicate the matrix structure in the two double orbital indices,
\begin{equation}
\hat{W}^{(r)} (q) = \hat\epsilon_r^{-1}(q) \hat{V} (\vec q) 
\end{equation}
with the screening function
\begin{equation}
\label{ }
\hat\epsilon_r (q) = \left[ \hat{\mathbf{1}}- \hat{V}(\vec q) \hat{\chi}^{(r)} (q) \right] \, .
\end{equation}
This formula is represented diagrammatically on the left side of Fig. \ref{5diags}. It includes the constrained polarization transformed into the orbital frame,
\begin{equation} 
\label{ }
\hat{\chi}^{(r)}_{(o_1o_2)(o_3o_4)}  (q) = 2 \frac{T}{N} \sum_{k \atop b_1,b_2 } 
u_{o_1b_1} (\vec k + \vec q) u^*_{o_2b_2} (\vec k) 
u_{o_4b_2} (\vec k) u^*_{o_3b_1} (\vec k+ \vec q) 
G_{b_1} (k+q ) G_{b_2} (k)\, [1-  \Pi_t(b_1,b_2)] \, ,
\end{equation}
in which we have indicated by the round brackets like $(o_1o_2)$ which orbital indices belong to the same vertex side. The $u_{ob}(\vec{k})$ are the matrix elements of the orbital-to-band transformation which diagonalizes the kinetic matrix $H_{oo'}(\vec{k})$ in the bare Hamiltonian (\ref{hamilton}).
The projector $\Pi_t(b_1,b_2)$ is unity if bands $b_1$ and $b_2$ are both target bands and zero otherwise. The prefactor 2 comes from the spin sum. 
In contrast with the full RG, for which the interaction depends on three wavevectors and frequencies, the  cRPA just modulates the transfer-wavevector $\vec{q}$ and transfer-frequency $\omega$ dependence of the interaction.

\section{Multiorbital truncated-unity functional fRG}
Over the last two decades, functional renormalization group methods have been applied to the Hubbard-like fermion lattice model in many different forms and with different degree of approximations\cite{metzner,platt}. The standard renormalization group formalism based on the Wetterich equation\cite{wetterich} for the generating functional for one-particle irreducible vertex functions prescribes
flow equations as function of a flowing energy scale, here called $\Lambda$. In this work we focus on the one-particle irreducible two-particle interaction vertex with orbital indices and ignore all higher vertices as well as the flow of the one-particle vertex, i.e. the selfenergy.  

\begin{figure} 
  \includegraphics[width=.48\columnwidth]{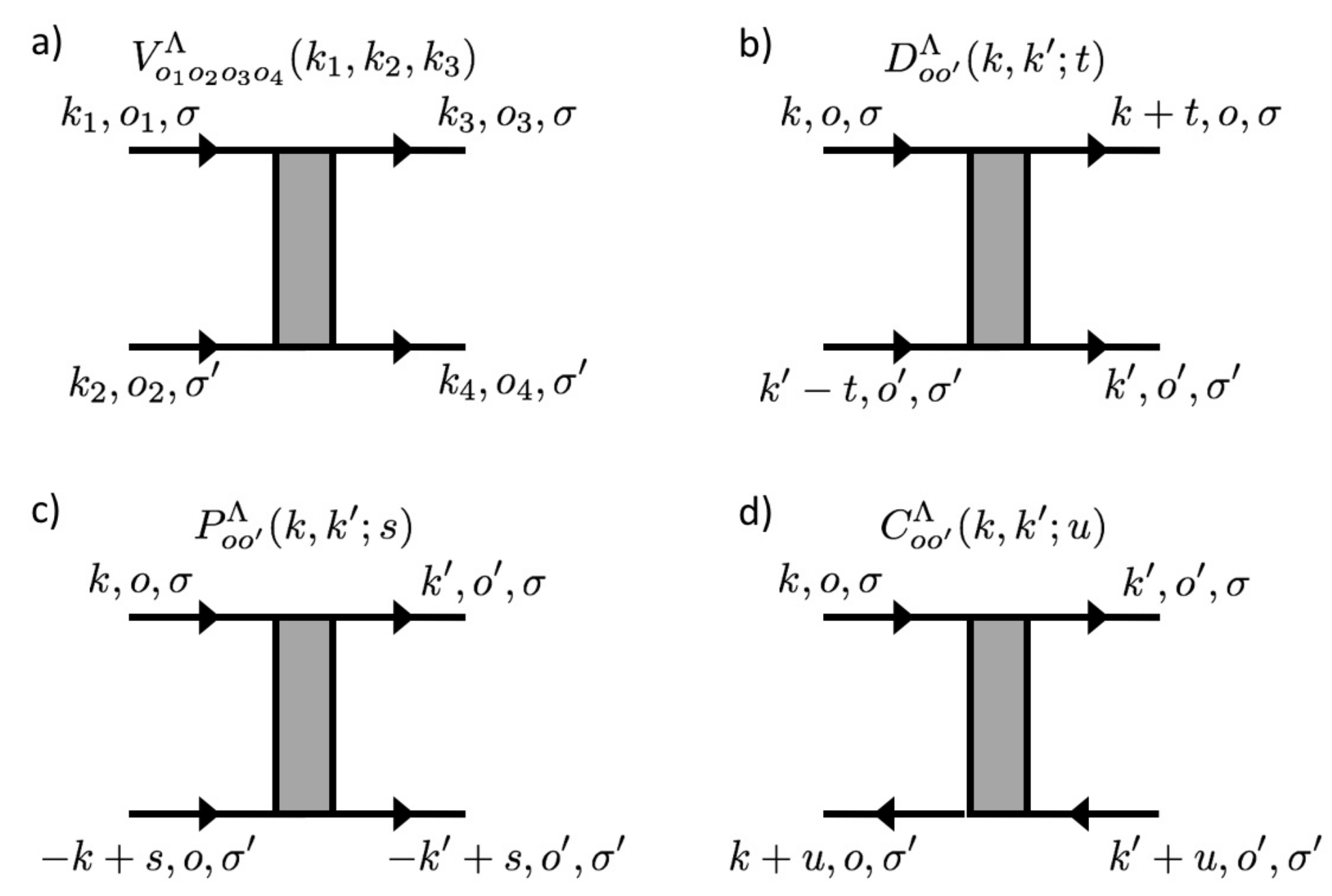}
 \caption{a) Coupling function for the spin-rotational multi-orbital case. b)- d) The three channels kept in the IOBI approximation of Subsec. \ref{iobi}. }
  \label{VPCD}
\end{figure}

Let us first describe the quantum numbers on which the interaction vertex depends on. We work with combined Matsubara-frequency/wavevector variables $k_i=(k_{0,i},\vec{k}_i)$ where a fermionic Matsubara frequency $k_{0,i}$ is an odd multiple of $\pi T$ and $\vec{k}_i$ is a wavevector in the first Brillouin zone. In addition we have the orbital indices $o_i$ which label the localized Wannier states that are as basis functions for the electronic states. Next we assume spin-rotational SU(2) invariance . Then the two-particle vertex can be described by a coupling function $V_{o_1o_2o_3o_4}^\Lambda (k_1,k_2,k_3)$. In this notation\cite{hsfr} (see also Fig. \ref{VPCD} a)), the two incoming particles $k_1,o_1$ and $k_2,o_2$ of the interaction carry spin projection $\sigma$ and $\sigma'$ and $k_3,o_3$ denotes the first outgoing particle with the same spin projection as the first incoming particle with $k_1,o_1$, i.e. $\sigma$. 
The momentum and frequency dependence of two-particle scattering can also be described in terms of Mandelstam variables 
\begin{equation}
\label{ }
s= k_1+k_2 \, , \quad t=k_3-k_1 \, , \quad \mbox{and} \quad u= k_4-k_1 \, , 
\end{equation}
with $k_4=k_1+k_2-k_3$ (modulo lattice).

\subsection{Channel decomposition and flow of vertex}
Based on the models introduced above we make the following channel-decomposed ansatz\cite{husemann,wang,metzner,lichtenstein} for the effective interaction at the initial energy scale $\Lambda_0$,
\begin{equation}
\label{ }
S^{\Lambda_0}_I = S^{\Lambda_0}_D + S^{\Lambda_0}_C + S^{\Lambda_0}_P 
\end{equation}
with 
\begin{equation}
\label{D0}
S^{\Lambda_0}_D  =\frac{T}{2N} \sum_{k,k',q,\sigma,\sigma' \atop o_1,o_2,o_3,o_4} D_{o_1o_2o_3o_4}^{\Lambda_0} (k,k';q) 
\left[ \bar{c}_{k+q,o_3,\sigma}    {c}_{k,o_1,\sigma} \right] \left[ \bar{c}_{k',o_4,\sigma'} {c}_{k'+q,o_2,\sigma'} \right] \, ,
\end{equation}
in the density-density channel,
 \begin{eqnarray} \nonumber
S^{\Lambda_0}_C  &=& \frac{T}{2N} \sum_{k,k',q,\sigma,\sigma' \atop o_1,o_2,o_3,o_4} C_{o_1o_2o_3o_4}^{\Lambda_0} (k,k';q)   
\bar{c}_{k',o_3,\sigma}  \bar{c}_{k+q,o_4,\sigma'} {c}_{k'+q,o_2,\sigma'}    {c}_{k,o_1,\sigma} 
\\ &=& -
 \frac{T}{2N} \sum_{k,k',q,\sigma,\sigma' \atop o_1,o_2,o_3,o_4} C_{o_1o_2o_3o_4}^{\Lambda_0} (k,k';q)   
\left[ \bar{c}_{k',o_3,\sigma}  {c}_{k'+q,o_2,\sigma'}  \right] \left[  \bar{c}_{k+q,o_4,\sigma'} {c}_{k,o_1,\sigma}  \right] \, , \label{C0}
\end{eqnarray}
in the spin-flip channel, and 
 \begin{equation}
\label{P0}
S^{\Lambda_0}_P  = 
\frac{T}{2N} \sum_{k,k',q,\sigma,\sigma' \atop o_1,o_2,o_3,o_4} P_{o_1o_2o_3o_4}^{\Lambda_0} (k,k';q)   \left[ \bar{c}_{k,o_3,\sigma} \bar{c}_{-k+q,o_4,\sigma'} \right] \left[ {c}_{-k'+q,o_2,\sigma'}  {c}_{k,o_1,\sigma} \right] \, 
\end{equation}
in the pair channel. By the square brackets we group the single fermion fields into fermion bilinears. In the D-channel, we have equal-spin bilinears, which can be further collected, by summing over the spin and $k$, to charge bilinears. In the C-channel, spin-flips from $\sigma$ to $\sigma'$ are possible, and spin bilinears are obtained upon summation over $k$ and spin, with appropriate Pauli matrices inserted in the spin summations. In the P-channel, we orbital pair-hopping or pair bilinears.
Below we discuss how to distribute the bare interactions discussed in Subsec. \ref{multiorbV} over these three channels.

We can also write this interaction in terms of the coupling function that contains all channels,
\begin{equation}
\label{SI0}
S^{\Lambda_0}_I = \frac{T}{2N} \sum_{k_1,k_2,k_3} V^{\Lambda_0}_{o_1o_2o_3o_4} (k_1,k_2,k_3)  \bar{c}_{k_3,o_3,\sigma} \bar{c}_{k_4,o',\sigma'} {c}_{k_2,o_2,\sigma'}  {c}_{k_1,o_1,\sigma}
\end{equation}
with $k_4=k_1+k_2-k_3$ modulo reciprocal lattice. Then
\begin{equation}
\label{V0}
V^{\Lambda_0}_{o_1o_2o_3o_4} (k_1,k_2,k_3)  = D_{o_1o_2o_3o_4}^{\Lambda_0} (k_1,k_4;k_3-k_1)  +C_{o_1o_2o_3o_4}^{\Lambda_0} (k_1,k_3;k_3-k_2) +P_{o_1o_2o_3o_4}^{\Lambda_0} (k_1,k_3;k_1+k_2)   \, .
\end{equation}
The channel-decomposed fRG takes advantage of the observation that the wavevector and (in parts) also the frequency dependence of the renormalized interaction usually can be understood as a sum of three different channels, which are again of density, spin and pair type. All of these three channels describe the interaction of a particular type of fermion bilinear and depend most strongly on a particular collective wavevector $q$ that upon Fourier transformation to real space or time is related to the space or time distance between the bilinears that interact. 

The flow equation in the common level-2 truncation\cite{metzner} for 
$V_{o_1o_2o_3o_4}^\Lambda (k_1,k_2,k_3)$ reads
\begin{eqnarray}
\frac{d}{d\Lambda} V_{o_1o_2o_3o_4}^\Lambda (k_1,k_2,k_3) =    \partial_\Lambda {P}^\Lambda_{o_1o_2o_3o_4}  (k_1,k_3;s)   + \partial_\Lambda {D}^\Lambda_{o_1o_2o_3o_4}  (k_1,k_4;t) + \partial_\Lambda {C}^\Lambda_{o_1o_2o_3o_4}  (k_1,k_3;u) 
\label{vdot} \end{eqnarray} 
with the one-loop particle-particle contributions $\partial_\Lambda {P}^\Lambda_{o_1o_2o_3o_4}  (k_1,k_3;s)$ and the two different particle-hole channels $\partial_\Lambda {D}^\Lambda_{o_1o_2o_3o_4}  (k_1,k_4;t)$ and $\partial_\Lambda {C}^\Lambda_{o_1o_2o_3o_4}  (k_1,k_3;u)$, where
\begin{eqnarray} \label{pdot} \partial_\Lambda
 {P}^\Lambda_{o_1o_2o_3o_4}  (k_1,k_3;s)  & = & \frac{T}{N} \sum_{k \atop o_5,o_6,o_7,o_8} V^\Lambda_{o_1o_2o_5o_6} (k_1,-k_1+s,k) \, 
\partial_\Lambda \left[ {G}^{\Lambda}_{o_5o_7} (k) {G}^{\Lambda}_{o_6o_8}  (-k+s) \right] V^\Lambda_{o_7o_8o_3o_4} (k,-k+s,k_3) 
 \\ \label{ddot} 
\partial_\Lambda  {D}^\Lambda_{o_1o_2o_3o_4}  (k_1,k_4;t) & = & -2 \frac{T}{N} \sum_{k \atop o_5,o_6,o_7,o_8} V^\Lambda_{o_1o_6o_3o_5} (k_1,k+t,k_1+t) \, 
\partial_\Lambda \left[ {G}^{\Lambda}_{o_5o_7} (k) {G}^{\Lambda}_{o_8o_6} (k+t) \right] V^\Lambda_{o_2o_7o_4o_8} (k,k_4+t,k+t) 
\\ \nonumber && +
\frac{T}{N} \sum_{k \atop o_5,o_6,o_7,o_8} V^\Lambda_{o_1o_6o_3o_5} (k_1,k+t,k_1+t)  \, 
\partial_\Lambda \left[ {G}^{\Lambda}_{o_5o_7} (k) {G}^{\Lambda}_{o_8o_6} (k+t)  \right] V^\Lambda_{o_2o_7o_8o_4} (k_4+t,k,k+t) 
\\ \nonumber  &&  +
\frac{T}{N} \sum_{k \atop o_5,o_6,o_7,o_8} V^\Lambda_{o_1o_6o_5o_3} (k+t,k_1,k_1+t) \, 
\partial_\Lambda \left[ {G}^{\Lambda}_{o_5o_7} (k) {G}^{\Lambda}_{o_8o_6} (k+t)   \right] V^\Lambda_{o_2o_7o_4o_8} (k,k_4+t,k+t)
 \\ \label{cdot}
 \partial_\Lambda   {C}^\Lambda_{o_1o_2o_3o_4}  (k_1,k_3;u) & = & \frac{T}{N} \sum_k V^\Lambda_{o_1o_6o_5o_4} (k_1,k+u,k) \, 
\partial_\Lambda \left[ {G}^{\Lambda}_{o_5o_6} (k) {G}^{\Lambda}_{o_8o_6} (k+u) \right] V^\Lambda_{o_2o_7o_8o_3} (k,k_3+u,k_3) 
\end{eqnarray}
In these equations, $N$ is the number of unit cells in the system which should be sent to infinity to convert the momentum sums to integrals.
The product of the two internal lines in the one-loop diagrams contains the single-particle Green's function  $G^{\Lambda}_{oo'} (k) = \sum_{b} R_\Lambda (b,k) u_{ob} (k) u_{o'b}^*(k) / \left[ - i \omega + \epsilon_b(\vec{k}) + R_\Lambda (b,k) \Sigma_\Lambda (k) \right]$ at RG scale $\Lambda$. The sum goes over the bands $b$ of the model.  In the full fRG formalism, the Green's functions $G^{\Lambda}_{oo'} (k)$ should contain selfenergy corrections. In this work, we ignore selfenergy corrections in the integration over the high-energy bands. In general, as this integration does not involve small energy denominators in loop diagrams, the selfenergy corrections should be dominated by Hartree-Fock contributions that would mainly lead to a deformed band structure. In order to understand the effects on the effective interactions we think that is even useful to suppress these band deformation effects for the time being. Currently, the selfenergy feedback is being explored actively in the fRG community and can be added to the cfRG in future works.

In the cfRG\cite{honerkampcfRG} with the momentum- or energy-shell cutoff function or regulator ${R}_\Lambda (b,k) $ suppresses the modes with $|\epsilon_b (\vec{k}) | > \Lambda$.
Its $\Lambda$-derivative $\dot {R}_\Lambda (b,k)$ confines the modes  to a momentum shell with $|\epsilon (\vec{k}) | \approx \Lambda$. In this work we however use a simpler, flat cutoff
\begin{eqnarray*}
{R}_\Lambda (b,k) &=& \Lambda  \qquad \mbox{for} \quad b\in \quad \mbox{high-energy bands,}  \\
{R}_\Lambda (b,k) &=& 1  \qquad \mbox{for} \quad b\in \quad \mbox{target bands}.  
\end{eqnarray*}
The second line guarantees that no loops with two target-band lines contribute to the right hand side of the flow equation, as $\dot{R}_\Lambda (b,k)=0$ for $b$ in the target band. The flow goes from $\Lambda =1$ down to $\Lambda =0$. This adiabatically removes the high-energy bands from the theory. As these bands are away from the Fermi surface, no divergences occur in the flow. Furthermore, different choices for the cutoff function should yield quite similar end results.

\begin{figure} 
\includegraphics[width=.8\columnwidth]{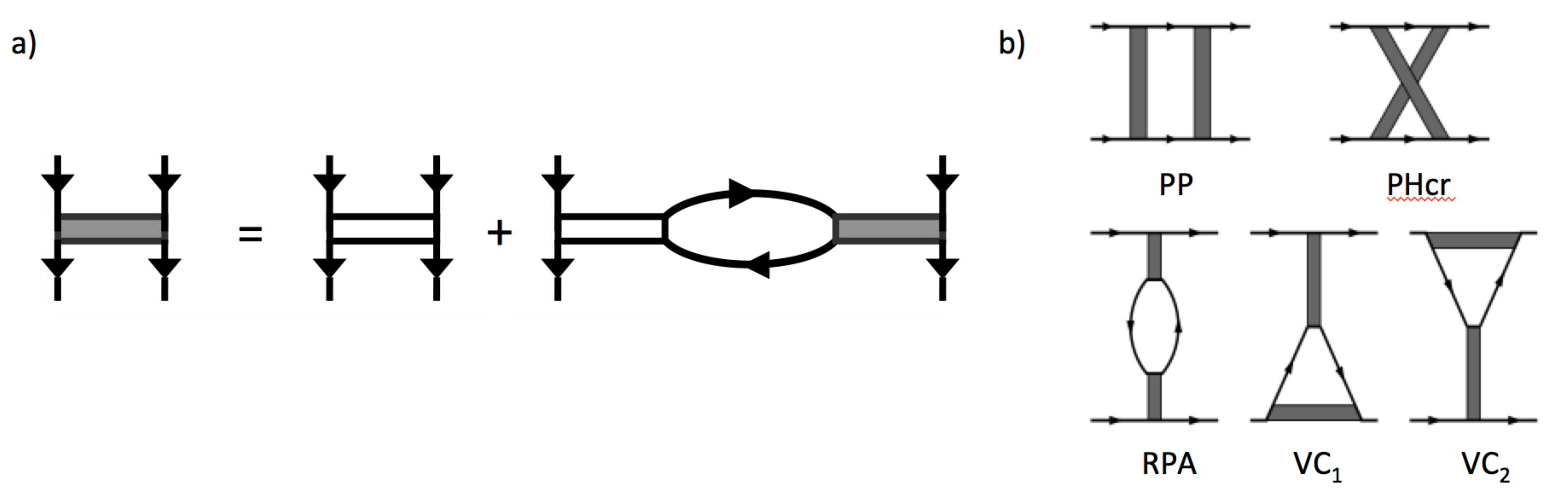}
 \caption{a) Effective cRPA interaction. The empty rectangles denote the bare  interaction and the full ones the screened interaction. The spin projection is conserved along the short edge of the rectangle (assuming spin-rotational invariance).  b) Five one-loop diagrams on the right hand side of the fRG flow equation for the interaction vertex. cfRG takes into account all diagrams. The internal lines carry cutoff functions and get differentiated with respect to the RG flow parameter in the flow equation. Only taking into account the diagram denoted as RPA amounts to cRPA.}
  \label{5diags}
\end{figure}

The diagrams for the right hand side are also shown inf Fig. \ref{5diags} b). As explained Ref. \onlinecite{honerkampcfRG}, the cRPA is obtained by only keeping the first term with the -2 in front on the right hand side of \ref{ddot}. 

At intermediate scales we have the following interaction in the effective action,
\begin{equation}
\label{ }
S^{\Lambda}_I = S^{\Lambda}_D + S^{\Lambda}_C + S^{\Lambda}_P 
\end{equation}
with 
\begin{equation}
\label{SD}
S^{\Lambda}_D  =\frac{T}{2N} \sum_{k,k',q,\sigma,\sigma' \atop o_1,o_2,o_3,o_4} D_{o_1o_2o_3o_4}^{\Lambda} (k,k';q) 
\bar{c}_{k+q,o_3,\sigma}  \bar{c}_{k',o_4,\sigma'} {c}_{k'+q,o_2,\sigma'}  {c}_{k,o_1,\sigma} \, ,
\end{equation}
in the density-density channel,
 \begin{equation}
\label{ }
S^{\Lambda}_C  =\frac{T}{2N} \sum_{k,k',q,\sigma,\sigma' \atop o_1,o_2,o_3,o_4} C_{o_1o_2o_3o_4}^{\Lambda} (k,k';q)   
\bar{c}_{k',o_3,\sigma} \bar{c}_{k+q,o_4,\sigma'} {c}_{k'+q,o_2,\sigma'}  {c}_{k,o_1,\sigma} \, ,
\end{equation}
in the spin-flip channel, and 
 \begin{equation}
\label{ }
S^{\Lambda}_P  = 
\frac{T}{2N} \sum_{k,k',q,\sigma,\sigma' \atop o_1,o_2,o_3,o_4} P_{o_1o_2o_3o_4}^{\Lambda} (k,k';q)   \bar{c}_{k,o_3,\sigma} \bar{c}_{-k+q,o_4,\sigma'} {c}_{-k'+q,o_2,\sigma'}  {c}_{k,o_1,\sigma} \, 
\end{equation}
in the pair channel.

\subsection{Form-factor expansion} \label{ffesec}
In the channel-decomposed fRG the $k$, $k'$-dependence of the three channels  $P_{o_1o_2o_3o_4}^{\Lambda} (k,k';q)$, $C_{o_1o_2o_3o_4}^{\Lambda} (k,k';q)$, and $D_{o_1o_2o_3o_4}^{\Lambda} (k,k';q)$ besides the dependence on the collective variable $q$ is captured via a form-factor expansion. In the literature (e.g. Ref.s \onlinecite{husemann,wang,lichtenstein}), this has so far been done for the wavevector dependence and not for the frequency dependence. Nevertheless, for what follows it is advantageous to lay out the basic formalism for expansions of both.

Hence we write, e.g. for the D-channel,
\begin{equation}
\label{ffe_or}
D_{o_1o_2o_3o_4}^{\Lambda} (k,k';q)  = \sum_{l,l'} \tilde D_{o_1o_2o_3o_4}^{\Lambda} (l,l';q) \tilde f_l(k) \tilde f^*_{l'}(k') \, ,
\end{equation}
where the $\tilde f_l(k)$ form an orthogonal basis system over the Brillouin zone of the lattice and over the Matsubara frequency axis,
\begin{equation}
\sum_{k} \tilde f^*_l(k) \tilde f_{l'}(k) = \delta_{\vec{l},\vec{l}'} \delta ( \tau- \tau') \, .
 \end{equation} 
Here we distinguish between spatial components $\vec{l}$ of the form-factor index $l$ and the Matsubara time $\tau \in [0,\beta]$ with $\beta =1/T$. 
The inversion of (\ref{ffe_or}) is then 
\begin{equation}
\label{ffproj_or}
\tilde D_{o_1o_2o_3o_4}^{\Lambda} (l,l';q)  =  \sum_{k,k'} D^{\Lambda}_{o_1o_2o_3o_4}  (k,k';q)    \tilde f_l^*(k)  \tilde f_{l'} (k') 
\end{equation}
A usual approximation for a discrete set of form-factor indices then consists in truncating the form-factor expansion after a certain expansion order $l_{\mathrm{max}} $ such that channel propagators $\tilde D_{o_1o_2o_3o_4}^{\Lambda} (l,l';q)$ become $l_{\mathrm{max}}\times l_{\mathrm{max}} $-matrices for each $q$. We can understand the physical meaning of the expansion (\ref{ffe_or}) by reinserting it into (\ref{SD}). This leads to an interaction of fermion bilinears that can be transformed back onto the real lattice and to Matsubara time,
\begin{equation}
\label{bilin}
 \sum_k \tilde f_l(k) \bar{c}_{k+q,o_3,\sigma} {c}_{k,o_1,\sigma} 
= \sum_{\vec{R},\vec{R}'} \int_0^\beta d\tau \int_0^\beta  d\tau' \, 
\bar{c}_{\vec{R},o_3,\sigma}(\tau)  {c}_{\vec{R}',o_1,\sigma} (\tau') \times
\frac{T}{N} \sum_{k_0,\vec{k}} \tilde f_l(k) e^{ik_0 (\tau-\tau') } e^{-i\vec{k} (\vec{R}-\vec{R}') } \, . 
\end{equation}
Now, a very physical route is to associate bond vectors $\vec{b}$ of the Bravais lattice with the spatial part of  $l$ and write $\tilde f_l(k)=\tilde f_{\vec{b}} (\vec{k}) f_{l_0} (k_0)$ with $\tilde f_{\vec{b}} (\vec{k}) = \sqrt{1 \over N} e^{-i\vec{k}\vec{b}}$. Then the wavevector sum in the end of the above expression gives $\delta_{\vec{b}, \vec{R}-\vec{R}'}$. Thus the bilinear lives on the bond $\vec{b}$, i.e. its creation and annihilation operators are by the vector $\vec{b}$ apart. If the main physics is given by short-bond bilinears, we can truncate the expansion (\ref{ffe_or}) for low $l$ or $\vec{b}$. 

The dependence on $k_0$ will determine the $\tau - \tau'$ structure of the bilinear in (\ref{bilin}). Now we use the simplest approximation and only keep $k_0$-independent frequency-form factors $\tilde f_0(k_0) =\sqrt{1/\beta} $, i.e. only consider $\tilde f_l (k) = \sqrt{1/\beta} \tilde f_{\vec{b}} (\vec{k})$. Then the bilinears on the right hand side above will be at equal times, as we obtain a factor $\delta (\tau- \tau')$ from (\ref{bilin}) when we sum over $k_0$. We can now introduce the {\bf instantaneous onsite approximation}, which uses exactly those frequency-independent form factors and ignores all bilinears on bonds of finite length $\vec{b} \not= 0 $. While it is not obvious that the instantaneous onsite approximation is a good approximation to the full flowing interaction in the general case, we note that the bare interactions are of this type, and that the usual single-channel summations like RPA or ladder summations for site-centered spin and charge fermion bilinears are kept in this approximation. The interaction is still allowed to develop a retardation and distance dependence between two bilinears, only the bilinears themselves remain local and instantaneous.    

For a reason explained below it is useful to rescale the form factors by $\sqrt{T/N}$ and introduce 
\begin{equation} f_l(k) = \sqrt{N \over T} \tilde f_l(k)  \, .
\end{equation}
For instance, using plane waves this would mean $ f_l(k) = e^{-i\vec{k}\vec{b}} e^{i k_0\tau}$. This choice leads to the orthogonality relation
\begin{equation}
\frac{T}{N} \sum_{k}  f^*_l(k) f_{l'}(k) =  \delta_{\vec{l},\vec{l}'} \delta ( \tau- \tau') \, 
 \end{equation} 
and the resolution of unity
 \begin{equation}
\frac{T}{N} \sum_{l}  f_l(k) f_{l}(k') =   \delta_{k,k'}  \, .
 \end{equation} 
With the rescaled form factors, using $D_{o_1o_2o_3o_4}^{\Lambda} (l,l';q)  = \frac{T}{N} \tilde D_{o_1o_2o_3o_4}^{\Lambda} (l,l';q)$, we have the expansion
\begin{equation}
\label{ffe}
D_{o_1o_2o_3o_4}^{\Lambda} (k,k';q)  = \sum_{l,l'} D_{o_1o_2o_3o_4}^{\Lambda} (l,l';q) f_l(k)  f^*_{l'}(k') \, ,
\end{equation}
and the projection
\begin{equation}
\label{ffproj}
 D_{o_1o_2o_3o_4}^{\Lambda} (l,l';q)  = \frac{T^2}{N^2} \sum_{k,k'}D^{\Lambda}_{o_1o_2o_3o_4}  (k,k';q)    f_l^*(k)  f_{l'} (k')  \, . 
\end{equation}
Here we see the advantage of the form-factor rescaling: if $D^{\Lambda}_{o_1o_2o_3o_4}  (k,k';q)$ is just a constant $D_0$, also the expansion coefficient $D_{o_1o_2o_3o_4}^{\Lambda} (0,0;q) $ will be $D_0$, without any additional $T$- and $N$-dependent prefactors.

With these conventions, the charge interaction then becomes in this instantaneous approximation 
\begin{equation} 
\label{SDio}
S^{\Lambda}_D  =\frac{T}{2N} \sum_{k,k',q,\sigma,\sigma' \atop o_1,o_2,o_3,o_4} D_{o_1o_2o_3o_4}^{\Lambda} (q) 
\bar{c}_{k+q,o_3,\sigma}  \bar{c}_{k',o_4,\sigma'} {c}_{k'+q,o_2,\sigma'}  {c}_{k,o_1,\sigma} \, .
\end{equation}
Here the $l=0$ indices in $\tilde D_{o_1o_2o_3o_4}^{\Lambda} (q)= \tilde D_{o_1o_2o_3o_4}^{\Lambda} (0,0;q)$ have not been written. We see from this formula that the interaction between charge bilinears $\bar{c}_{k+q,o_3,\sigma} {c}_{k,o_1,\sigma}$ and $\bar{c}_{k',o_4,\sigma'} {c}_{k'+q,o_2,\sigma'}$ can still become long-ranged in space and Matsubara time, provided that the $q$-dependence develops sufficient structure in $\tilde D_{o_1o_2o_3o_4}^{\Lambda} (q)$.
Furthermore, to make the structure clearer it is also useful to stow away  the orbital dependence (here $o_1,o_3$ for one bilinear and $o_2,o_4$ for the other bilinear) in combined indices $L=(l,o_1,o_3)$ and $L'=(l',o_2,o_4)$. Then the expansion reads (dropping the $\tilde \ $s again)
\begin{equation}
\label{ffeL}
D_{o_1o_2o_3o_4}^{\Lambda} (k,k';q)  = \sum_{L,L'}D^{\Lambda} (L,L';q) 
f_L(k, o_1,o_3) f^*_{L'}(k',o_2,o_4) \, ,
\end{equation}
where the form factors depend on the $l^{(')}$-content of $L^{(')}$ times indicator functions for specific orbital combinations,
\begin{equation}
\label{ }
f_{L=(l,o,\tilde{o})} (k,o_1,o_3) = f_l(k) \delta_{o\tilde{o},o_1o_3} \, .
\end{equation}
With these, we can also define general channel-projections of any interaction function $V_{o_1o_2o_3o_4}^{\Lambda} (k,k',k+q)$ as
\begin{equation}
\label{ }
\hat{D}_{L,L'} [ V^\Lambda ] (q) = \frac{T^2}{N^2} \sum_{k,k',o,\tilde{o},o',\tilde{o}'} 
V_{o,o',\tilde{o},\tilde{o}'}^{\Lambda} (k,k',k+q) f^*_L (k,o, \tilde{o})  f_{L'} (k',o',\tilde{o}') \, .
\end{equation}
In the same vein, we define projected particle-particle and particle-hole loop matrices
\begin{equation}
\label{ }
\hat{L}^\Lambda_{\mathrm{PP},L,L'}  (q) =\frac{T}{N}\sum_{k,o,\tilde{o},o',\tilde{o}'} f^*_L(k,o,\tilde{o}) G_{o,o'} (k) G_{\tilde{o},\tilde{o}'} (-k+q) f_L(k,o',\tilde{o}') 
\end{equation}
and
\begin{equation}
\label{ }
\hat{L}^\Lambda_{\mathrm{PH},L,L'}  (q) =\frac{T}{N}\sum_{k,o,\tilde{o},o',\tilde{o}'} f^*_L(k,o,\tilde{o}) G_{o,o'} (k) G_{\tilde{o},\tilde{o}'} (k+q) f_L(k,o',\tilde{o}') \, .
\end{equation} 
\subsection{Truncated-unity fRG equations}
In the so-called truncated-unity fRG (TUfRG), one uses the above-described channel decomposition. Furthermore one employs the form factor expansion of the three channels. This provides three sets of one-loop flow equations, one for $D^{\Lambda} (L,L';q)= \hat{D}^{\Lambda}_{L,L'} (q) $, one for $C^{\Lambda} (L,L';q)= \hat{C}^{\Lambda}_{L,L'} (q)$ and one for $P^{\Lambda} (L,L';q)= \hat{P}^{\Lambda}_{L,L'} (q)$. 
On the right hand sides one always has full vertices as the sum over the three channels plus the initial interactions. 

By inserting truncated resolutions of unity in the form factors of the type
\begin{equation}
\delta_{k,k'} = \sum_{L} f_L (k) f_L^* (k') \, ,
\end{equation}
one can cast these equations in the form of TUfRG matrix equations\cite{lichtenstein},
\begin{eqnarray}
\label{dP}
\frac{d}{d\Lambda} \hat{P}^\Lambda (q) &=& \hat{P} [ V^\Lambda] (q)  \dot{\hat{L}}^\Lambda_{\mathrm{PP}}  (q) \hat{P} [ V^\Lambda] (q)  \\
\frac{d}{d\Lambda} \hat{C}^\Lambda (q) &=& \hat{C} [ V^\Lambda] (q)  \dot{\hat{L}}^\Lambda_{\mathrm{PH}}  (q) \hat{C} [ V^\Lambda] (q) \label{dC}
\end{eqnarray}
and 
\begin{equation}
\label{dD}
\frac{d}{d\Lambda} \hat{D}^\Lambda (q) = -2 \hat{D} [ V^\Lambda] (q)  \hat{L}^\Lambda_{\mathrm{PH}}  (q) \dot{\hat{D}} [ V^\Lambda] 
(q) + 
\hat{C} [ V^\Lambda] (q)  \dot{\hat{L}}^\Lambda_{\mathrm{PH}}  (q) \hat{D} [ V^\Lambda] (q) +
\hat{D} [ V^\Lambda] (q) \dot{ \hat{L}}^\Lambda_{\mathrm{PH}}  (q) \hat{C} [ V^\Lambda] (q) \, . 
\end{equation}
These equations can be integrated numerically. In a previous work\cite{lichtenstein} we have shown how the projections of $V^\Lambda$ can be done on the real lattice in an efficient way and that the TUfRG scheme exhibits a good scalability on parallel computers. Nevertheless, for what follows below these tweaks are not really needed, as the integration over high-energy bands requires much less momentum resolution.  

Alternatively, it may appear advantageous to keep the form factors as simple properties of  the Bravais lattice and not to lump together form-factor and orbital indices. Then the TUfRG equation still carry the orbital indices explicitly, with matrices in the form-factor indices only,
\begin{eqnarray}
\label{dPo}
\frac{d}{d\Lambda} \hat{P}^\Lambda_{o_1o_2o_3o_4}  (q) &=& \sum_{o_5,o_6,o_7,o_8} \hat{P} [ V_{o_1o_2o_5o_6}^\Lambda] (q)  \dot{\hat{L}}^\Lambda_{\mathrm{PP},{o_5o_6o_7o_8}}  (q) \hat{P} [ V_{o_7o_8o_3o_4}^\Lambda] (q)  \\
\frac{d}{d\Lambda} \hat{C}_{o_1o_2o_3o_4}^\Lambda (q) &=& \sum_{o_5,o_6,o_7,o_8} \hat{C} [ V_{o_1o_8o_5o_4}^\Lambda] (q)  \dot{\hat{L}}^\Lambda_{\mathrm{PH},{o_5o_6o_7o_8}}  (q) \hat{C} [ V_{o_6o_2o_3o_7}^\Lambda] (q) \label{dCo}
\end{eqnarray}
and 
\begin{eqnarray}
\frac{d}{d\Lambda} \hat{D}_{o_1o_2o_3o_4}^\Lambda (q) &=& \sum_{o_5,o_6,o_7,o_8} \left \{ -2 \hat{D} [ V_{o_1o_8o_3o_5}^\Lambda] (q)  \dot{\hat{L}}^\Lambda_{\mathrm{PH},{o_5o_6o_7o_8}}  (q) \hat{D} [ V_{o_2o_6o_4o_7}^\Lambda] 
(q)\right.  \nonumber \\ && + 
\hat{C} [ V_{o_1o_8o_5o_3}^\Lambda] (q)  \dot{\hat{L}}^\Lambda_{\mathrm{PH}},{o_5o_6o_7o_8}  (q) \hat{D} [ V_{o_2o_6o_4o_7}^\Lambda] (q) \nonumber \\ && + \left. 
\hat{D} [ V_{o_1o_8o_3o_5}^\Lambda] (q)  \dot{\hat{L}}^\Lambda_{\mathrm{PH},{o_5o_6o_7o_8}}  (q) \hat{C} [ V_{o_2o_6o_7o_4}^\Lambda] (q) \right\} \, ,  \label{dDo}
\end{eqnarray}
with
\begin{equation}
\hat{L}^\Lambda_{\mathrm{PP},{o_5o_6o_7o_8}} (q)  = \frac{T}{N} \sum_k G^{\Lambda}_{o_5o_7} (k) G^{\Lambda}_{o_6o_8} (-k+q)
\end{equation}
and
\begin{equation}
\hat{L}^\Lambda_{\mathrm{PH},{o_5o_6o_7o_8}}  (q) = \frac{T}{N} \sum_k G^{\Lambda}_{o_5o_6} (k) G^{\Lambda}_{o_7o_8} (k+q) \, .
\end{equation}

\section{Approximation strategies}
The fRG equations (\ref{dP}), (\ref{dC}) and (\ref{dD}) or (\ref{dPo}), (\ref{dCo}) and (\ref{dDo}) can in principle be solved in general for the full matrix structure with three channel couplings $ \hat{P}^\Lambda$,  $\hat{C}^\Lambda$ and $\hat{D}^\Lambda$  each dependening on four orbital indices and a number of form factors. In the one-orbital case has been done e.g. in Ref. \onlinecite{lichtenstein} for the one-band Hubbard model with a larger number of form factors. In particular, the convergence of the results with respect to the length of the form-factor expansion could be demonstrated. However, for a many-band problem the overall effort grows considerably, as the number of couplings grows with the fourth power of the number of Wannier orbitals considered. Thus, additional approximations appear useful.
We start with an approximation that reduces the complexity of the  frequency and momentum  dependence. After that we simplify the orbital structure.

\subsection{Instantaneous bilinears and static channel coupling} \label{staticCC}
The TUfRG flow equation above were derived for simultaneous form-factor expansions in wavevector and frequency dependence. We already defined the the instantaneous onsite  approximation in Subsec. \ref{ffesec}, where only the zero-bondlength and equal-time bilinears are kept, equivalent to constant from factors in wavevector and frequency. 
In the TUfRG equations above, (\ref{dPo}), (\ref{dCo}) and (\ref{dDo}), this would require to project the channel couplings on the instantaneous contribution, by a 'flat' (i.e., with constant form factor) summation over all frequencies. This creates the difficulty of treating the high-frequency behavior of the with sufficient precision, which makes the approach at least more complex and potentially numerically challenging.      

Hence,  for this work, we will not use frequency-projected couplings and loop diagrams and only used the TUfRG-scheme for wavevector dependence of the right hand side of the flow equations. We will still restrain the interaction to be between instantaneous bilinears, but use a different approach to deal with the frequency summation on the right hand side of Eqs.   (\ref{dPo}), (\ref{dCo}) and (\ref{dDo}). This approach was laid out previously in the context of one-dimensional\cite{cla,markhof} and two-dimensional one-band models\cite{reckling}. There it was argued that one-frequency parametrizations with only the collective Mandelstam frequency kept\cite{karrasch} provide a good approximation. This is indeed equivalent to the instantaneous bilinear approximation, but there are no frequency-form-factor projections used in the flow equations. To clarify the approximation, 
we restrict the form-factor expansion and projection to the wavevector content, leading to quantities with $\hat{}$ -hats like $\hat{P}^\Lambda (k_0,k_0';q)$,  and keep the dependence on two fermionic (in front of the semicolon) and one collective bosonic Matsubara frequency (behind the semicolon in $q=(q_0,\vec{q})$) explicit. Then the flow equation for the pair-channel coupling  would read (suppressing the orbital indices for the moment, these are the same as in Eq. \ref{dPo})
\begin{eqnarray}
\frac{d}{d\Lambda} \hat{P}^\Lambda (k_0,k_0';q) &=& T \sum_{k_0''} \hat{P}^\Lambda 
\left[ P^\Lambda (k,k'';q) + C^\Lambda ( k,-k'';k+k''-q) +D^\Lambda (k,-k''+q;k''-k) \right] \nonumber \\  && \qquad \cdot   \dot{\hat{L}}^\Lambda_{\mathrm{PP}} (k_0'',q) 
\hat{P}^\Lambda \cdot 
\left[ P^\Lambda (k'',k';q) + C^\Lambda ( k'',-k';k'+k''-q) +D^\Lambda (k'',-k'+q;k'-k'') \right] \, .
\end{eqnarray}
Here, the loop cannot be summed separately over the Matsubara frequency $k_0''$ as the couplings also depend on $k_0''$. We have more precisely, with $k=(k_0,\vec{k})$,
\begin{equation}
\hat{L}^\Lambda_{\mathrm{PP}} (k_0,q)  = \frac{1}{N} \sum_{\vec{k}} G^{\Lambda} (k) G^{\Lambda} (-k+q) \, .
\end{equation} 
We notice that the summation frequency $k_0''$ appears in the third, collective-frequency argument of the for the pair channel 'non-native' channel couplings. Hence, in the sum over $k_0''$ we might have to sum over a sharp peak, if the collective-frequency dependence of any of these couplings becomes sharp.  This makes the implementation more costly. In Refs. \onlinecite{cla,markhof,reckling} it is argued that at least for the cases studied it makes sense to pull the couplings out of the $k_0$ sum by setting the collective frequency in the non-native couplings to a certain frequency $\bar{q}$. This chosen to be 0 in \onlinecite{cla,markhof} and  0 or the RG scale in Ref. \onlinecite{reckling}.

Here we also use this approximation and we write
\begin{equation}
\frac{d}{d\Lambda} \hat{P}^\Lambda (q) = \hat{P}^\Lambda 
\left[ P^\Lambda (q) + C^\Lambda (\bar{q}) +D^\Lambda (\bar{q}) \right]  T \sum_{k_0''}  \dot{\hat{L}}^\Lambda_{\mathrm{PP}} (k_o'',q) 
\hat{P}^\Lambda 
\left[ P^\Lambda (q) + C^\Lambda ( \bar{q} ) +D^\Lambda (\bar{q}) \right] \, .
\end{equation}
In the applications below we choose $\bar{q}=0$. This can be called the static channel-coupling approximation. In recent literature\cite{cla,markhof}, this choice is also named 'coupled-ladder approximation'. For the orbital content, we use the scheme described in Eqs. \ref{dPo}, \ref{dCo}, and \ref{dDo}.

\subsection{Onsite approximation} \label{onsite}
As next approximation, we truncate the form-factor expansion of Eq. \ref{ffe} quite early and only keep the unit-cell-local form factor, which is constant in the Brillouin zone and frequency independent, along the discussion after Eq. \ref{bilin}. Then the projections of Eq. \ref{ffproj} become simply Brillouin zone averages over the momentum,
\begin{equation}
\label{ff00proj}
D_{0,0} \left[ V_{o_1o_2o_3o_4}^{\Lambda} (k,k';q) \right]  = \frac{1}{N^2} \sum_{\vec k,\vec k'} V_{o_1o_2o_3o_4}^{\Lambda} (\vec k,\vec k'; q)   
\end{equation}
The question is of course whether this quite drastic simplification is justifiable. We argue that the bare interactions described in Subsec. \ref{multiorbV} reflect this unit-cell local or onsite structure, as in the two-center approximation for the Coulomb matrix elements it only consists of interacting onsite charge, spin and pair bilinears.  Most importantly, the standard Kanamori parametrization in terms of local interaction parameters $U$ and $U'$ for the intra- and inter-orbital onsite repulsions and $J$ for the Hund's coupling can also be considered as falling into this class. 

It is clear that the onsite approximation misses some important physics that could come in the form interactions between non-local bilinears. For instance, $d$-wave pairing on the square lattice and most other unconventional Cooper pairing examples involve pair bilinears on neighbored sites and in neighboring unit cells. One might however suspect that this physics, unless it is already present in the bare interaction, will only become relevant in the solution of the low-energy effective model. This is not aimed for here, as we focus on integrating out the bands away from the Fermi level.
Furthermore, in systems where the orbitals are centered not at the same but at various places in the unit cell, the restriction of bilinears formed between orbitals only inside the same unit cell may become questionable. As an example, consider a chain of alternating A and B orbitals centered next to each other along the chain. Each unit cell contains one A orbital and one B orbital. Then the nearest-neighbor density-density interaction will have two parts, one where onsite density bilinears with operators $a_i^\dagger a_i$  couple to $b_i^\dagger b_i$ in the same unit cell and one, where they couple to $b_{i-1}^\dagger b_{i-1}$ in a neighboring unit cell. In the D-channel, this non-locality in the interaction between onsite bilinears is no problem and well-captured by the $\vec{q}$-dependence of the channel coupling. However, the projection with unit-cell-local form factor only on other channels , $C$ or $P$, will miss the one half of this interaction that extends out of the unit cell. In such cases, the truncation has to be adapted to take into account all physically equivalent interactions.
  
\subsection{Intraorbital bilinear approximation}  \label{iobi}
With the approximations of the last two subsections\ref{staticCC} and \ref{onsite}, the flowing multi-orbital interactions arising from the bare interactions discussed in Subsec. \ref{multiorbV} all involve onsite charge, spin or pair bilinears formed by two fermionic operators at the same lattice site, albeit with possibly different orbital index. At the initial scale $\Lambda_0$, we can write
\begin{eqnarray}
D_{o_1o_2o_3o_4}^{\Lambda_0} (l,l';q) &=& 
U \delta_{l,0}\delta_{l',0} \delta_{o_1,o}\delta_{o_3,o} \delta_{o_2,o}\delta_{o_4,o} 
+  U'  \delta_{l,0}\delta_{l',0} \delta_{o_1,o}\delta_{o_3,o} \delta_{o_2,o'}\delta_{o_4,o'} (1- \delta_{oo'})  \nonumber \\&&
+ V^{\rho_{oo'},\Lambda_0} (q)  \delta_{l,0}\delta_{l',0} \delta_{o_1,o}\delta_{o_3,o} \delta_{o_2,o'}\delta_{o_4,o'}  = D_{oo'}^{\Lambda_0} (q) \,  , \label{Dini3}  \\[1mm]
C_{o_1o_2o_3o_4}^{\Lambda_0} (l,l';q) &=& 
  J  \delta_{l,0}\delta_{l',0} \delta_{o_1,o}\delta_{o_4,o} \delta_{o_2,o'}\delta_{o_3,o'} (1- \delta_{oo'})  = C_{oo'}^{\Lambda_0} (q) \label{Cini3}
  \,  , \\[1mm]
P_{o_1o_2o_3o_4}^{\Lambda} (l,l';q) &=& 
  J  \delta_{l,0}\delta_{l',0} \delta_{o_1,o}\delta_{o_2,o} \delta_{o_3,o'}\delta_{o_4,o'} (1- \delta_{oo'})  = P_{oo'}^{\Lambda_0} (q) \, . \label{Pini3}
\end{eqnarray}
This distribution of the initial interaction is not unique, and below we will discuss and use an alternative choice.
By comparing with Eqs. (\ref{D0}), (\ref{C0}), and (\ref{P0}), we find that the initial interaction is made from intra-orbital bilinears, where both constituent fermions carry the same orbital index.

Now it appears to be a meaningful approximation to keep only interactions with orbital combinations corresponding to the terms appearing in the bare interaction, i.e. that involve for {\bf i}ntr{\bf a}{\bf o}rbital {\bf b}ilinears that {\bf i}nteract. We call this the IOBI (intra-orbital bilinear) approximation. 
The corresponding couplings are shown diagrammatically in Fig \ref{VPCD} b) to d). 
We will see below numerically that in many cases in the integration over high-energy bands, the flow of additional interorbital bilinears, which can also be kept in the more general truncation of Subsec. \ref{9c}, will be less important. 

These restrictions in the IOBI-approximation amount to writing
\begin{equation}
\label{ }
D_{o_1o_2o_3o_4}^{\Lambda} (l,l';q) = D_{oo'}^{\Lambda} (q) \delta_{l,0}\delta_{l',0} \delta_{o_1,o}\delta_{o_3,o} \delta_{o_2,o'}\delta_{o_4,o'} 
\end{equation}
in the direct and 
\begin{equation}
\label{ }
C_{o_1o_2o_3o_4}^{\Lambda} (l,l';q) = C_{oo'}^{\Lambda} (q) \delta_{l,0}\delta_{l',0} \delta_{o_1,o}\delta_{o_4,o} \delta_{o_2,o'}\delta_{o_3o'} 
\end{equation}
in the crossed channel. In the pair channel we only consider intraorbital onsite pairs (effectively excluding spin-triplet or any non-local pairs),
\begin{equation}
\label{ }
P_{o_1o_2o_3o_4}^{\Lambda} (l,l';q) = P_{oo'}^{\Lambda} (q) \delta_{l,0}\delta_{l',0} \delta_{o_1,o}\delta_{o_2,o} \delta_{o_3,o'}\delta_{o_4o'}\, . 
\end{equation}
This way, the orbital and wavevector structure of the $D_{oo'}^{\Lambda} (q)$-interaction fits naturally on the particle-hole bubble RPA diagram on the right hand side of the TUfRG equations. 
The internal bubble contains a summation over the orbital index.  
Similarly, the Hund's  rule spin interactions $C_{oo'}^{\Lambda} (q)$ match the crossed particle-hole channel on the right hand side of $\partial_\Lambda C^\Lambda$, and a ladder summation of these diagrams with $C_{oo'}^{\Lambda} (q)$ alone  would maintain its orbital structure. Finally, the pair hopping interaction $P_{oo'}^{\Lambda} (q)$ is kept naturally in the particle-particle channel.

In this IOBI-flow, no new bilinear interactions terms are generated. It is the distance- and frequency-dependence between the onsite charge-, spin- and pair bilinears that get altered.

After some RG steps, e.g. when the screening or non-target bands have been integrated out, it makes sense to study the resulting effective interaction. In the IOBI approximation this can be obtained as
\begin{equation}
\label{ }
V^\Lambda_{o_1o_2o_3,o_4} (k_1,k_2,k_3) = P_{o_1o_3}^{\Lambda} (k_1+k_2) +  D_{o_1o_2}^{\Lambda} (k_3-k_1) +  C_{o_1o_2}^{\Lambda} (k_3-k_2) \, .
\end{equation}

As laid out in the previous papers on the cfRG\cite{honerkampcfRG,kinza}, the cRPA would be obtained from the cfRG by only taking into account the RPA diagram in the D-channel  (see Fig. \ref{5diags} for the RPA diagram and the other diagrams). In the IOBI-approximation to the cRPA this would mean that only the charge interactions originating from the non-local Coulomb interaction and in the local intra- and interorbital density-density terms $U$ and $U'$ are renormalized. 
The Hund's rule spin and pair hopping interactions do not flow, as those terms are kept in the C- and P-channel, and these channels are not included in the cRPA flow.

\subsection{Two-orbital approximation} \label{9c}
\begin{figure} 
  \includegraphics[width=.69\columnwidth]{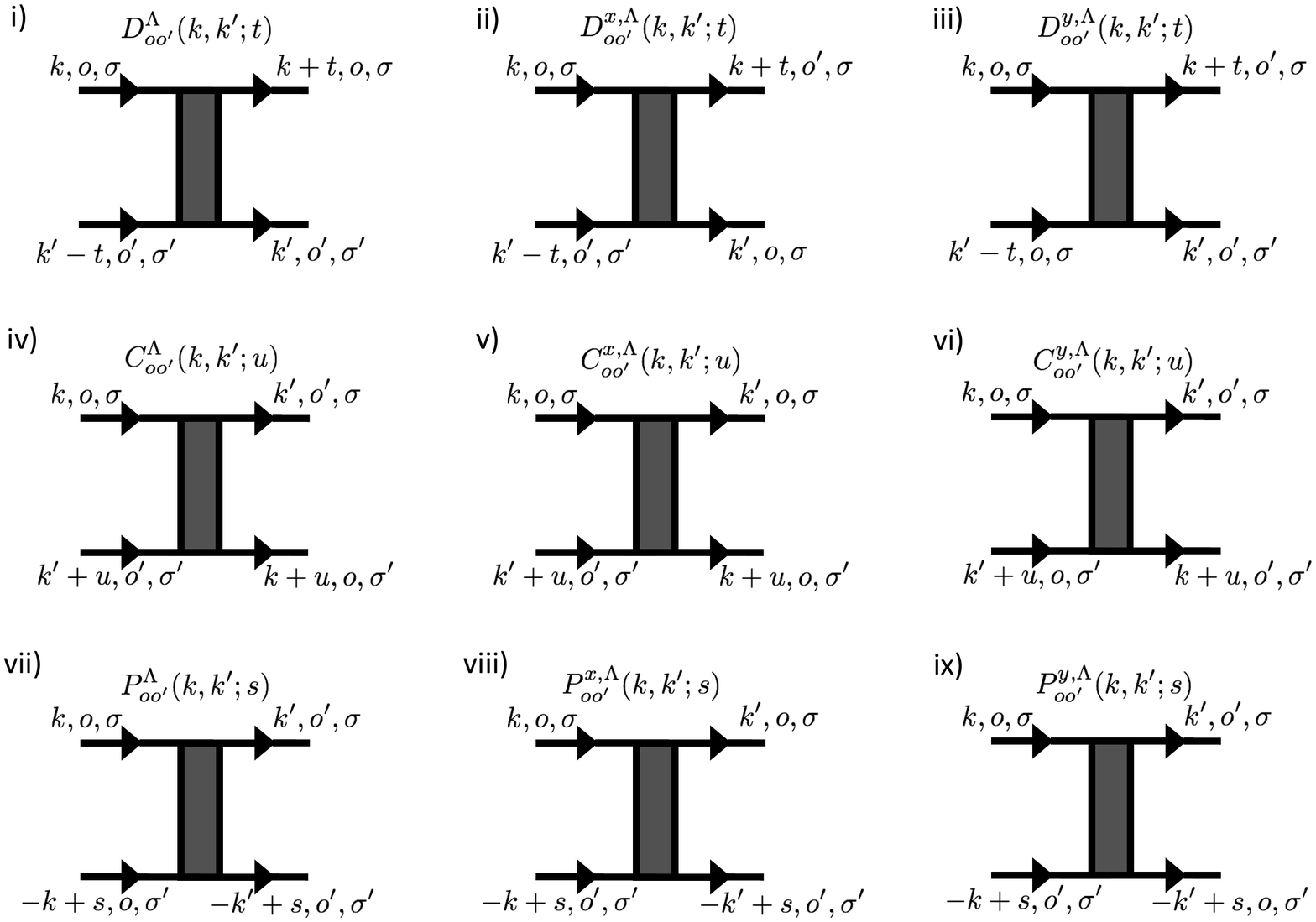}
 \caption{ The nine coupling functions kept in the two-orbital approximation of Subsec. \ref{9c}. The functions i), iv) and vii) are the ones already considered in the IOBI approximation of Subsec. \ref{iobi}.}
  \label{9csfig}
\end{figure}

An alternative truncation of the orbital content, still with only local form factors, which remedies the shortcoming of the cRPA flow to include the Hund's rule terms but also makes the study of two-orbital interactions more complete, is to consider the enlarged set of 9 effective interaction functions $D_{oo'}^{\Lambda} (q)$, $D_{oo'}^{x/y,\Lambda} (q)$ , $C_{oo'}^{\Lambda} (q)$, $C_{oo'}^{x/y,\Lambda} (q)$, $P_{oo'}^{\Lambda} (q)$, and $P_{oo'}^{x/y,\Lambda} (q)$, or more precisely,
\begin{eqnarray}
D_{o_1o_2o_3o_4}^{\Lambda} (l,l';q) & =&  \delta_{l,0}\delta_{l',0} \left[ 
D_{oo'}^{\Lambda} (q) \delta_{o_1,o}\delta_{o_3,o} \delta_{o_2,o'}\delta_{o_4,o'}  +
D_{oo'}^{x,\Lambda} (q) \delta_{o_1,o}\delta_{o_4,o} \delta_{o_2,o'}\delta_{o_3,o'}  +
D_{oo'}^{y,\Lambda} (q) \delta_{o_1,o}\delta_{o_2,o} \delta_{o_3,o'}\delta_{o_4,o'} \right]   \nonumber \\ \\ 
C_{o_1o_2o_3o_4}^{\Lambda} (l,l';q) &= &  \delta_{l,0}\delta_{l',0}  \left[  C_{oo'}^{\Lambda} (q)  \delta_{o_1,o}\delta_{o_4,o} \delta_{o_2,o'}\delta_{o_3o'} 
+C_{oo'}^{x,\Lambda} (q)  \delta_{o_1,o}\delta_{o_3,o} \delta_{o_2,o'}\delta_{o_4o'} 
+C_{oo'}^{y,\Lambda} (q)  \delta_{o_1,o}\delta_{o_2,o} \delta_{o_3,o'}\delta_{o_4o'} \right]  \\
P_{o_1o_2o_3o_4}^{\Lambda} (l,l';q) &=& 
\delta_{l,0}\delta_{l',0} \left[ 
P_{oo'}^{\Lambda} (q)  \delta_{o_1,o}\delta_{o_2,o} \delta_{o_3,o'}\delta_{o_4o'}
+ P_{oo'}^{x,\Lambda} (q)  \delta_{o_1,o}\delta_{o_3,o} \delta_{o_2,o'}\delta_{o_4o'}
+ P_{oo'}^{y,\Lambda} (q)  \delta_{o_1,o}\delta_{o_4,o} \delta_{o_2,o'}\delta_{o_3o'}  \right]
\, .
\end{eqnarray}
As all these interactions still only depend on two orbital  indices, we can call this the {\bf two-orbital approximation}. The coupling functions are shown graphically in Fig \ref{9csfig}. Now, for a specific channel D, C, or P with momentum-/frequency-transfer $q$ we allow for all three possible orbital combinations $o_1o_2o_3o_4$ with two pairs of identical orbital indices. The constituents of the bilinears that exchange this $q$ are not necessarily in the same orbital, but in the approximation used here with just the constant form factor, they are always in the same unit cell. 

The larger set of couplings gives more freedom to allows the Hund's rule terms to flow even if only the D-channel flow is considered, i.e. if one restrains the study to cRPA.
Namely, using the two-orbital approximation, we can now assign all terms of the bare interaction to the D-couplings: The density-density interactions go as before into $D_{oo'}^{\Lambda} (q)$, while the momentum-independent and non-retarded Hund's rule spin terms are put into $D_{oo'}^{x,\Lambda} (q)$ and the Hund's rule pair hopping terms are assigned to $D_{oo'}^{y, \Lambda} (q)$. The initial condition for the C- and P-couplings is then zero, but those couplings are generally generated during the flow.
We have checked numerically that the assignment of the Hund's rule 

Below we will see that in the cases studied numerically, the differences between the intra-orbital approximation and the more general two-orbital approximation are in may cases on the quantitative level.

\section{Effective target band interactions}
We can now employ either cRPA or the IOBI-fRG in order to integrate out the non-target bands. This yields an effective interaction $V^\Lambda_{o_1o_2o_3,o_4} (k_1,k_2,k_3)$ that can be projected onto the target band $b=t$ by 
\begin{equation}
\label{ }
V^\Lambda_{tttt} (k_1,k_2,k_3) = \sum_{o_1,o_2,o_3,o_4} u_{o_1t} (\vec{k}_1) u_{o_2t} (\vec{k}_2) u^*_{o_3t} (\vec{k}_3) u^*_{o_4t} (\vec{k}_4) V^\Lambda_{o_1o_2o_3,o_4} (k_1,k_2,k_3)
\end{equation}   
We can now define wavevector-resolved static charge and spin couplings as those couplings that appear in the bubble summation for the RPA charge susceptibility or in the crossed ladder summation for the RPA spin susceptibility. For the  coupling to non-retarded charge source fields, the appropriate vertex summed in the RPA loop diagrams over the frequency arguments of the first incoming ($k_1$) and second outgoing line ($k_4$) with a fixed transfer frequency $\omega$ between the first incoming ($k_1$) and first outgoing line ($k_3$). In the spin channel, the frequency transfer is between the second incoming ($k_2$) and and first outgoing line ($k_3$), and the summation is over  $k_1$ and $k_3$. 

For the vertex in the target band, we use the channel couplings obtained by the cfRG flow. More precisely, we use for the charge coupling in the IOBI approximation with momentum transfer $\vec{q}$ and frequency transfer $\omega$
\begin{eqnarray}
V^{\Lambda}_{tttt,\mathrm{ch}} (\vec{k}_1,\vec{k}_2,\vec{k}_3=\vec{k}_1+\vec q; \omega) & =& \sum_{o,o'} \left[  u_{o,t} (\vec{k}_1) u_{o',t} (\vec{k}_2) u^*_{o,t} (\vec{k}_1+\vec{q} ) u^*_{o',t} (\vec{k}_2- \vec{q}) D_{oo'}^{\Lambda} (\vec q, \omega) \right.  \nonumber \\&&
+  u_{o,t} (\vec{k}_1) u_{o',t} (\vec{k}_2) u^*_{o',t} (\vec{k}_1+\vec{q} ) u^*_{o,t} (\vec{k}_2- \vec{q}) C_{oo'}^{\Lambda} (\vec{k}_2-\vec{q}-\vec{k}_1;  \tilde\omega)  \nonumber \\ &&\left. 
+  u_{o,t} (\vec{k}_1) u_{o,t} (\vec{k}_2) u^*_{o',t} (\vec{k}_1+\vec{q} ) u^*_{o',t} (\vec{k}_2- \vec{q}) P_{oo'}^{\Lambda} (\vec{k}_1+\vec{k}_2;\tilde\omega) \right]  \, ,
\label{Veffch}
\end{eqnarray}
and for the spin coupling
\begin{eqnarray}
V^{\Lambda}_{tttt,\mathrm{sp}} (\vec{k}_1,\vec{k}_2,\vec{k}_3=\vec{k}_2+\vec q; \omega) & =& \sum_{o,o'} \left[  u_{o,t} (\vec{k}_1) u_{o',t} (\vec{k}_2+\vec{q}) u^*_{o,t} (\vec{k}_2+  \vec{q} ) u^*_{o',t} (\vec{k}_1- \vec{q}) D_{oo'}^{\Lambda} (\vec{k}_2+\vec{q}+\vec{k}_1, \tilde\omega)  \right. \nonumber \\&&
+  u_{o,t} (\vec{k}_1) u_{o',t} (\vec{k}_2) u^*_{o',t} (\vec{k}_2+ \vec{q} ) u^*_{o,t} (\vec{k}_1- \vec{q}) C_{oo'}^{\Lambda} (\vec{q}, \omega )  \nonumber \\ && \left. 
+  u_{o,t} (\vec{k}_1) u_{o,t} (\vec{k}_2) u^*_{o',t} (\vec{k}_2+\vec{q} ) u^*_{o',t} (\vec{k}_1- \vec{q}) P_{oo'}^{\Lambda} (\vec{k}_1+\vec{k}_2,\tilde\omega) \right]   \, .
\label{Veffsp}
\end{eqnarray}
Here the subscript $tttt$ indicates that all legs of the vertex are in the target band.
The main contribution to the frequency-dependence of the charge channel comes from the 'channel-native'  coupling  $D_{oo'}^{\Lambda} (\vec{q},\tilde \omega )$ and in the spin channel from the there-native coupling $C_{oo'}^{\Lambda} (\vec{q},\tilde \omega )$. For the contributions by the non-native couplings, i.e. $C_{oo'}^{\Lambda} (\vec{q},\tilde \omega )$ and $P_{oo'}^{\Lambda} (\vec{q},\tilde \omega )$ 
in the case of the charge coupling and 
$D_{oo'}^{\Lambda} (\vec{q},\tilde \omega )$ and $P_{oo'}^{\Lambda} (\vec{q},\tilde \omega )$ 
in the case of the spin coupling,
the variables $\tilde \omega$ are placeholders for combinations of external leg frequencies and the transfer frequency. This means that the loop sums average over the frequency argument $\tilde \omega$. The question is then which frequency $\tilde \omega$ we should choose here as representative (without having to do the full target band RPA).
We could set $\tilde \omega=0$ again. This would be consistent with the static-channel coupling approximation used for the flow of the channels (see Subsec. \ref{staticCC}). However, as also argued in Ref. \onlinecite{reckling}, the one-frequency parametrizations without form-factor expansion are designed to get the leading flow at low frequencies right and cannot be expected to give good results for higher frequencies of the external fermionic legs of the interaction. This can also be observed in the charge and spin couplings defined above. Here, using the static channel coupling approximation, the high-$\omega$ values of the effective interactions would not go back to the bare values, if the zero-frequency values for the non-native couplings are used.
This does not make sense, as all renormalizations are due to loop corrections which vanish for high frequencies. Similarly, using frequency-averages leads to the same problem in the high-frequency behavior.
A more suitable approximation is then to set $\tilde \omega = \omega$, which is done for (\ref{Veffch}) and (\ref{Veffsp}) in the data discussed below. This guarantees the correct disappearance of perturbative corrections in the high-$\omega$-limit and merges with the static-channel coupling approximation in the case of small $\omega$.   

For the more general two-orbital approximation, the expressions (\ref{Veffch}) and (\ref{Veffsp}) are generalized accordingly.
Next, in order to obtain expression that only depend on the transfer wavevector and frequency, we average over the two incoming wavevectors, which can again be understood as onsite-bilinear approximation, 
This leads to 
\begin{equation}
\label{ }
V^\Lambda_{t,\mathrm{ch}} (\vec q, \omega) =  \frac{1}{N^2}\sum_{\vec{k}_1,\vec{k}_2} V^\Lambda_{tttt,\mathrm{ch}} (\vec{k}_1,\vec{k}_2,\vec{k}_3=\vec{k}_1+\vec q; \omega) \, ,
\end{equation}
and
\begin{equation}
\label{ }
V^\Lambda_{t,\mathrm{sp}} (\vec q, \omega) =  \frac{1}{N^2}\sum_{\vec{k}_1,\vec{k}_2} V^\Lambda_{tttt,\mathrm{sp}} (\vec{k}_1,\vec{k}_2,\vec{k}_3=\vec{k}_2+\vec q;\omega) \, .
\end{equation}
These quantities can be compared between the cRPA and cfRG calculations for $V^\Lambda_{o_1o_2o_3,o_4} (k_1,k_2,k_3)$. For this comparison it is further useful to analyze the static dielectric charge screening function  
\begin{equation}
\label{ }
\epsilon_{t,\mathrm{ch}} (\vec q, \omega) = \frac{V^{\Lambda_0}_{t,\mathrm{ch}} (\vec q, \omega) }{V^\Lambda_{t,\mathrm{ch}} (\vec q,\omega) } \, , 
\end{equation}
which is the ratio between the bare charge interaction in the target band at RG scale $\Lambda_0$ and effective charge interaction in the target band after the screening bands have been integrated out at RG scale $\Lambda$. Similarly we define the spin screening function
\begin{equation}
\label{ }
\epsilon_{t,\mathrm{sp}} (\vec q, \omega) = \frac{V^{\Lambda_0}_{t,\mathrm{sp}} (\vec q,\omega) }{V^\Lambda_{t,\mathrm{sp}} (\vec q,\omega) } \, .
\end{equation}

\section{Results in three-band model}
Here we show results for the effective interactions in charge and spin channel as well as for the screening functions. We use the three-orbital model described in Sec. \ref{multiorbV} and use either cRPA or cfRG to integrate out the two bands below and above the Fermi level. Most data was obtained using an 18$\times$18 wavevector grid in the Brillouin zone and 18 positive bosonic Matsubara frequencies. These numbers can be increased significantly if needed, as the flow part of the code scales linearly in the numbers of wavevectors and frequencies.  

First, let us study the situation for the purely onsite initial interactions using the two approximations with three or nine coupling functions as in Subsecs. \ref{iobi} and \ref{9c}. 

\begin{figure}
\includegraphics[width=.48\columnwidth]{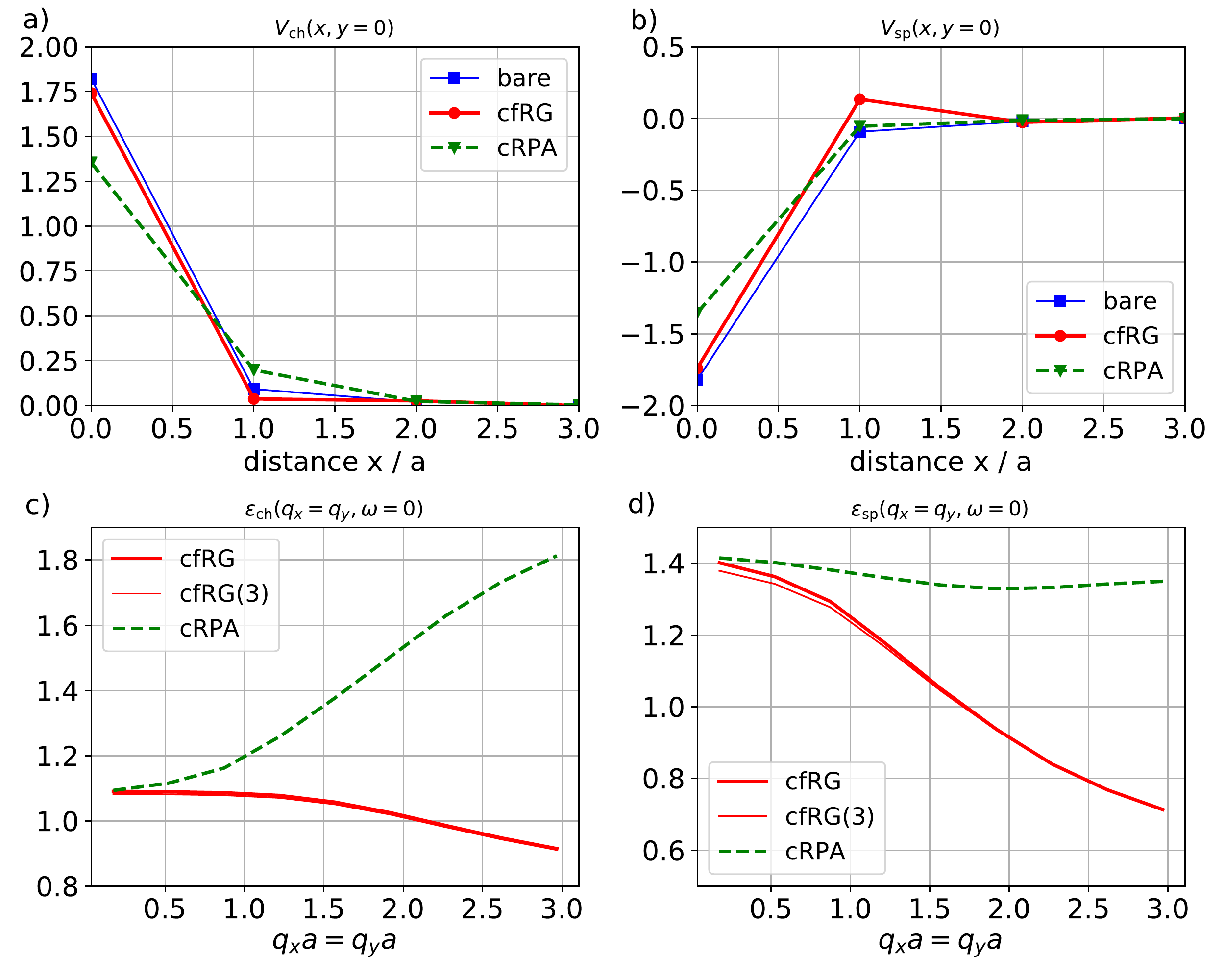}%
 \caption{Comparison of the effective target band interactions obtained with cRPA and cfRG with the bare target band interaction, for onsite bare interactions $U=6$, $U'=0$, $J=0$, $V_c=0$. The upper plots show the real space dependence of the effective interactions along the $x$-direction on the real lattice, for the charge channel in the left plot and for the spin channel in the right plot. Note that although the bare interaction in the three-orbital model is purely local, the orbital-to-band transformation makes the bare interaction within the central band slightly non-local already. The lower plots show the static  screening functions in the target band in the charge (left) and spin (right) channel as function of the wavevector along the Brillouin zone diagonal. The plots a) and b) in the upper half are for the two-orbital approximation of Subsec. \ref{9c}. In the lower plots c) and d) we distinguish between the  IOBI approximation, called cfRG(3),  with just three intra-orbital bilinears allowed and the more general two-orbital cfRG with in total 9 interactions channels. For the charge channel the curve lie on top of each other in this case.}
  \label{Uplot}
\end{figure}

In Fig. \ref{Uplot} we show data for pure intraorbital onsite interactions, with $U=6$, $U'=J=V_c=0$.  In the upper left plot we show the spatial dependence of the charge interaction for the bare case, the cRPA and the cfRG. The cfRG values in the upper plots are the ones from the two-orbital approximation of Subsec. \ref{9c}. We notice that even the bare interaction has acquired non-local contributions. This occurs because the orbital-to-band transformation is momentum-dependent. The onsite term is significantly reduced by the cRPA compared to the bare value, but only slightly by the cfRG. In contrast with this, the cRPA produces an enhanced nearest neighbor repulsion, while the cfRG lowers the nearest-neighbor repulsion compared to the bare interaction. Similar differences are observed for the effective spin interaction. Notably, the spin nearest-neighbor term in cfRG is repulsive, favoring antiferromagnetic alignment, while it is negative and smaller in magnitude in the bare and cRPA interactions. This generation of nonlocal AF couplings was already observed in Ref. \onlinecite{kinza}. It is possibly understood from the momentum structure of interband particle-hole bubbles, somewhat as a weaker reminiscence of how RKKY interactions come about in a metal. 

\begin{figure}
\includegraphics[width=.48\columnwidth]{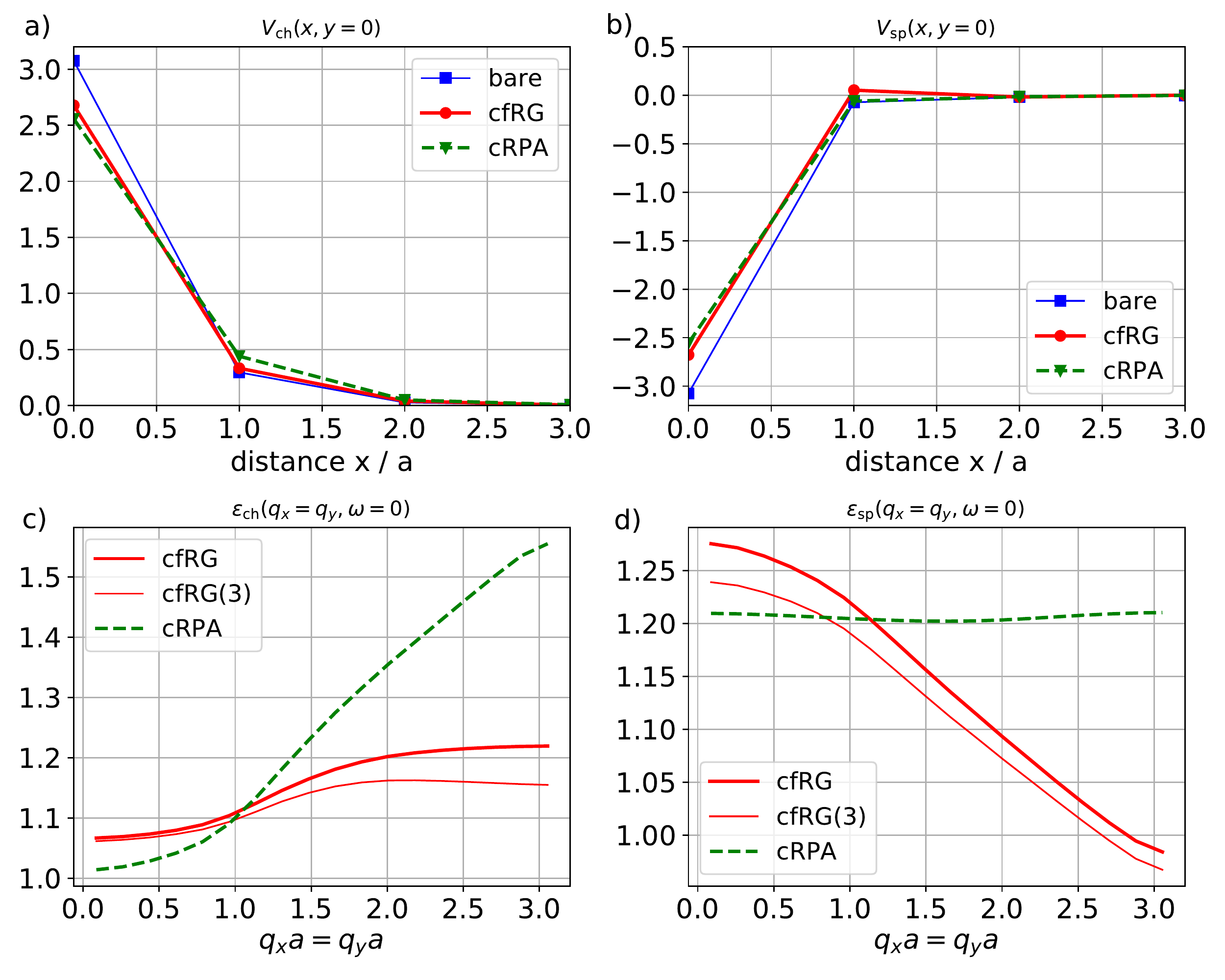}%
 \caption{Same as in Fig. \ref{Uplot}, but now for onsite Kanamori interactions $U=4.8$, $U'=3.6$, $J=0.6$, $V_c=0$. }
  \label{KaEGplot}
\end{figure}

Another specific choice of interaction parameters $U=4.8$, $U'=3.6$, with $J=0.6$ is shown in Fig. \ref{KaEGplot}. 
We see that now the onsite charge interaction in the cRPA is only slightly more reduced compared to the bare one than in cfRG. 
For the nearest-neighbor charge interaction the situation is again reversed, here the cRPA value is higher than the bare one and the cfRG one. Also for the spin interactions, the onsite terms and nearest neighbor terms are renormalized differently. The nearest neighbor spin term in cfRG is again positive.

In the lower panels of Figs. \ref{Uplot} and  \ref{KaEGplot} we plot the target-band screening functions in charge and spin channels. Now we also compare the data for the IOBI approximation with three coupling functions with that of the two-orbital approximation. For the pure intraorbital onsite repulsion in Fig. \ref{Uplot} the two-orbital approximation (thick lines) does not change the results of the IOBI (thin lines, if visible), while for the case in Fig. \ref{KaEGplot}, quantitative differences can be observed.
The different spatial variation of the effective interactions in cRPA and both cfRG versions leads to quite different momentum dependencies of the screening functions. Most notably, the charge screening in cRPA is far stronger than in cfRG for half of the Brillouin zone at larger $\vec{q}$, in line with the differences at small distances in the upper panel. Also, the spin screening function in the cfRG gets smaller than 1 for large $q_x = q_y  \to \pi$, i.e. there we have antiscreening or enhancement  of a staggered interaction component that can of course also be found in the distance dependence of the real-space spin interaction in the panel above.


\begin{figure}[t]
 \includegraphics[width=.48\columnwidth]{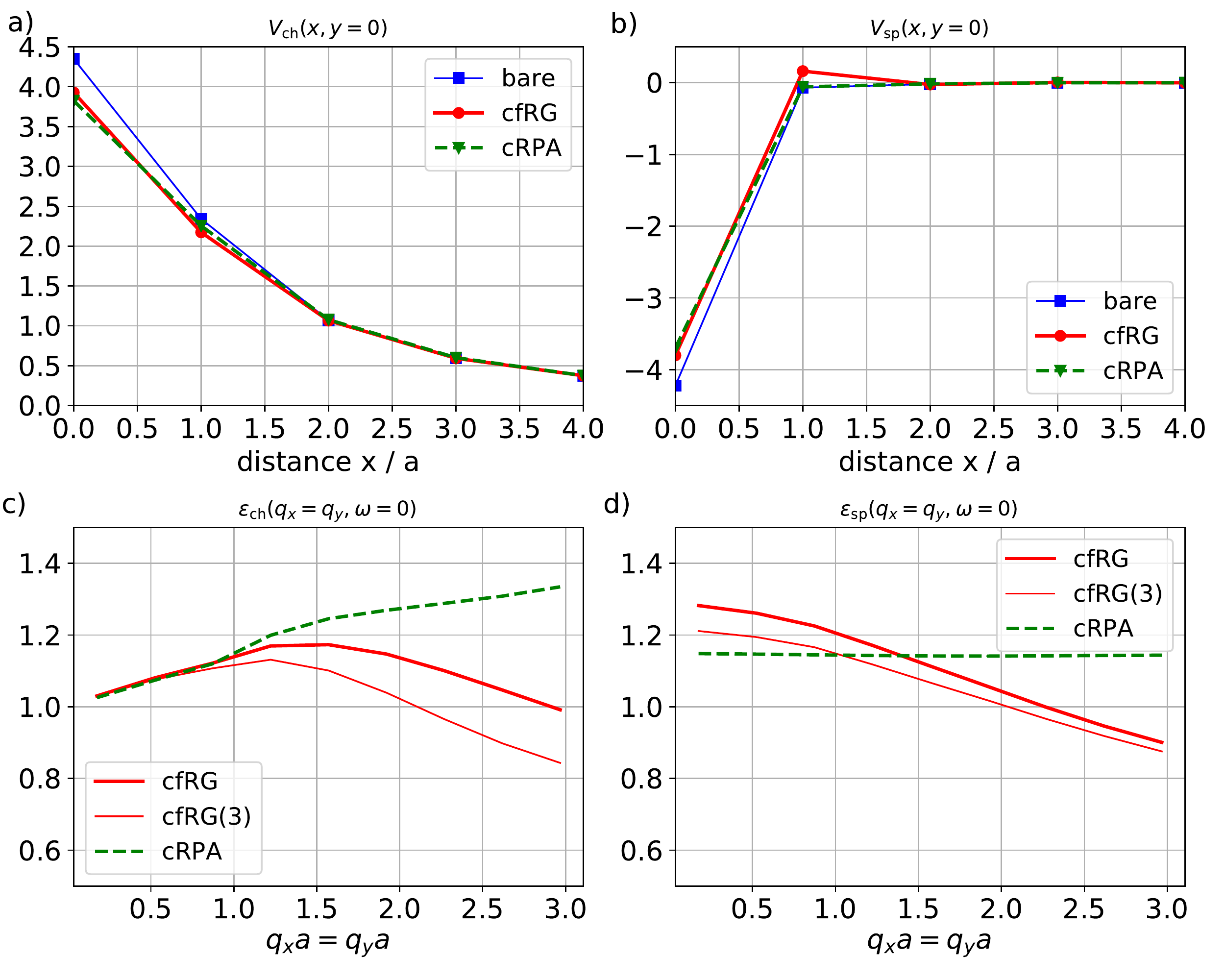}
 \caption{Comparison of the effective target band interactions obtained with cRPA and cfRG with the bare target band interaction, for nonlocal bare interactions with a screening length of 6 lattice constants, $U=6$ $U'=4.8$, $J=0.6$ and $V_c=2$. The upper plots show the real space dependence of the effective interactions along the $x$-direction, for the charge channel in the left plot and for the spin channel in the right plot. The lower plots show the static  screening functions in the charge (left) and spin (right) channel as function of the wavevector along the Brillouin zone diagonal. The plots a) and b) in the upper half are for the two-orbital approximation. In the lower plots c) and d)  distinguish between the IOBI approximation, called cfRG(3), with just three intra-orbital bilinears allowed and the more general two-orbital cfRG with in total 9 interactions channels.}
  \label{EGplot}
\end{figure}

\begin{figure}
 \includegraphics[width=.48\columnwidth]{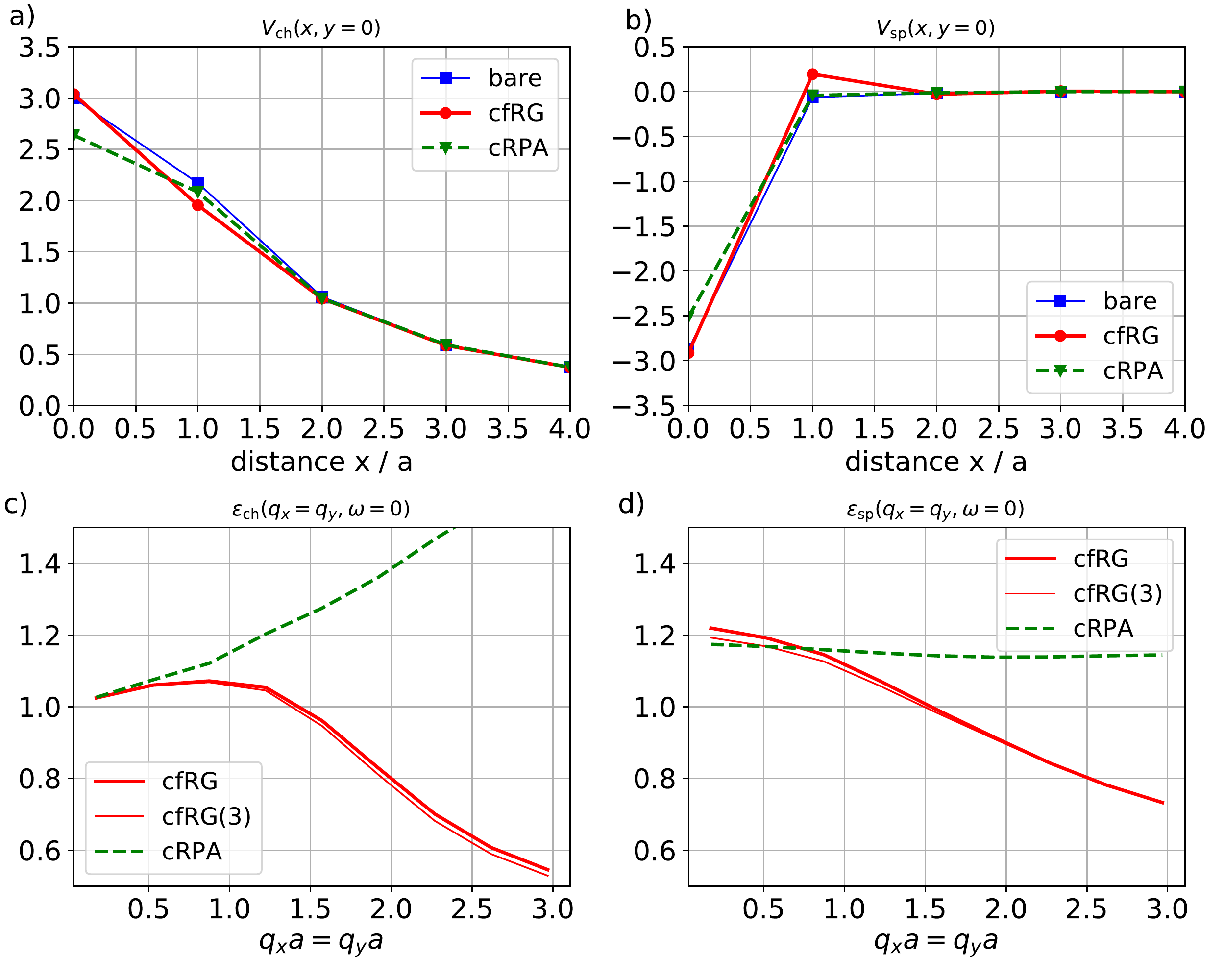}
 \caption{Same as in Fig \ref{EGplot}, but with $U=6$ $U'=2$, $J=0$ and $V_c=2$.}
  \label{EGplot2}
\end{figure}

Comparable and partially stronger differences at short distances and higher wavevectors are found when a longer-ranged bare interaction is used. In Figs. \ref{EGplot} and \ref{EGplot2}we show the same set of data for interaction parameters that include a density-density repulsion with decay length of 6 lattice sites. Now in the second data set, Fig. \ref{EGplot2}, the target-band onsite charge interaction in cfRG is even larger than the bare one, while the cRPA screens it down. For both cases, the spin interaction is positive on the nearest neighbor site again, leading to increased AF tendencies. The most striking difference is found in the dielectric or charge screening function for the case in Fig. \ref{EGplot2} where $\epsilon_{\mathrm{ch}} (\vec{q})$  now  becomes smaller than 1 in the cfRG for larger $q$  while is rises above 1 for the cRPA.  Hence there is significant anti-screening for wavevectors near $(\pi,\pi)$. Similarly to the parameter sets plotted above but even more strongly,  the spin screening function dives below 1 as well for $\vec{q} \to (\pi,\pi)$ in Fig. \ref{EGplot2}. 

Hence, the cfRG static screening properties can be quite different from those in cRPA, at least at small distances of the order of one lattice spacing.


\begin{figure}
  \includegraphics[width=.48\columnwidth]{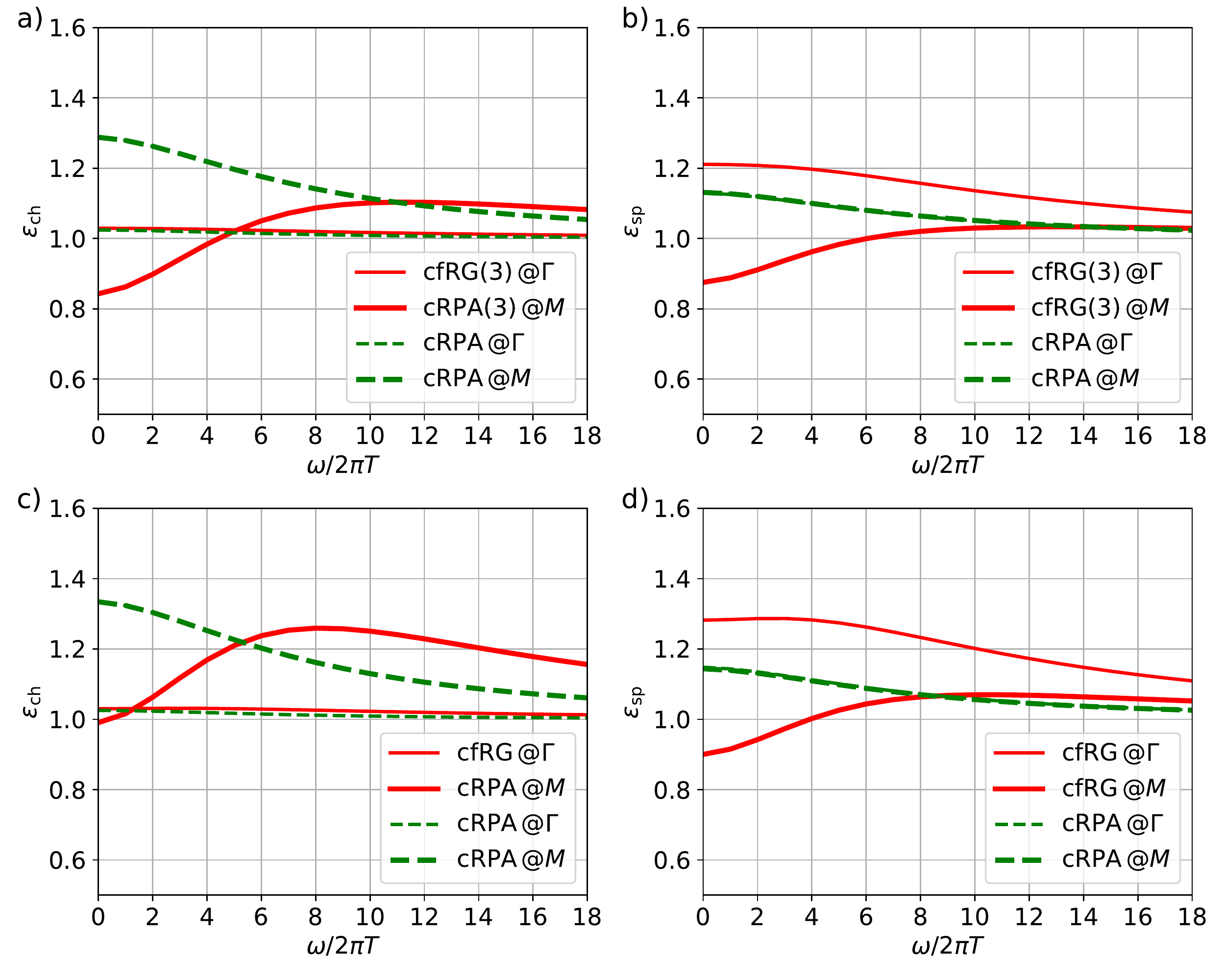}
 \caption{Comparison between cRPA and cfRG data for the Matsubara frequency dependence of the charge (left, a) and c)) and spin (right, b) and d)) screening functions with the interaction parameters for onsite Kanamori interactions $U=6$, $U'=4.8$,$J=0.6$, $V_c=3$ as well as a screening length of 6 lattice sites  at $T=0.1$, for two different wavevectors in the two-dimensional Brillouin zone. }
  \label{dynEGplot}
\end{figure}

This picture is complemented by the frequency-dependent screening for which we show some exemplary data in Fig. \ref{dynEGplot}, obtained with the one-frequency parametrization in the static channel-coupling approximation, as described in Subsec.\ref{staticCC}. Here the frequency dependence of the charge screening in cRPA and cfRG is quite different for wavevectors near $M=(\pi,\pi)$. The antiscreening of the cfRG persists far over a Matsubara frequency range of the order of the gap between the bands, which is 4 to 5 in our open energy units.  Near $\Gamma$ the differences are small for the charge screening. Also in the spin channels, cRPA and cfRG give different results, again with some antiscreening near $\vec{q}=(\pi,\pi)$.
In the lower panels of \ref{dynEGplot} we also show data obtained the more general two-orbital approximation described in Subsec.\ref{9c}. We notice some smaller differences compared  to the IOBI, which maintain the differences between cRPA and cfRG.  
All qualitative findings are the same compared to IOBI. Thus the two-orbital calculation should be more precise because it ignores less interactions, but is presumably not needed for most cases to a reasonable first estimate of the screening.  

\section{Discussion}
We conclude this presentation of approximation strategies in the cfRG and comparison data by summarizing that the two methods, cfRG and cRPA show indeed differences that in certain cases may be strong enough to lead to qualitatively different predictions when the screened interaction are employed in the solution of the low-energy model. For instance, the ratio between onsite and nearest-neighbor interactions often decides on the type of ground state ordering, or in other cases  strongly influences the energy scale for ordering. 
In particular this ratio was found to differ in cfRG and cRPA: the ratio differed by 20$\%$ between cRPA and cFRG for the parameters used in Fig. \ref{EGplot2}, and for bare intraorbital interaction only as in Fig. \ref{Uplot}, the effective nearest-neighbor repulsion in cfRG was only a fifth of that in cRPA. 
Furthermore, additional spin AF interactions that are generated in the effective model by the cfRG but not by the cRPA can play a role in cases of competing ordering tendencies. Finally, also the frequency dependencies of the screening functions was found to differ qualitatively for parts of the momentum space. 
We plan to extend our studies in order to include the instability analysis that can directly show the resulting changes in ground state order and energy scales for these to occur. 

We note  that in our three-band model study all screening effects are somewhat small. They are big enough to make the differences between different diagrammatic content of cRPA and cfRG visible. However our computed screening functions range between 0.6 and 1.5. So they do not deviate strongly from unity. This is in contrast with the screening values given in the literature, we can exceed 2 for instance in graphene systems\cite{roesner} or reach values $\sim 4-5$ in iron superconductors\cite{miyakeFe}. We hope that our new, numerically lighter IOBI or two-orbital approaches can be applied in a more realistic setting in order to understand, if these numerical differences are just caused by the use of too small bare interactions in our model, or if other factors increase the screening in real materials. Another aspect that should be understood is if the real space staggering tendencies found in the spin channel in the cfRG with the generation of a nearest neighbor AF component are somehow linked to band structure of the chosen model. The data shown here were for a target-band filling of $\sim  60\%$ which should not be a special point. Varying the band filling of the target band by $\pm15\%$  did not change the results much. 

On the methodical side, the IOBI or two-orbital approximations are two differently truncated approximations to the full orbital dependence of the effective interactions that both only keep terms that at most depend on two orbital indices. The usual bare interactions including Hund's rule terms can already be represented by the IOBI couplings that employs just three functions to represent the orbital structure, momentum and frequency structure of the interaction.
For most parameter sets, in particular those with $J=0$, the IOBI and the more general two-orbital approximation that uses nine functions give quite similar results. However, our data shows that for $J \not 0$ there are some quantitative corrections to the IOBI in the two-orbital approximation such that, if the numerical constraints permit, the two-orbital scheme may give more precise results. Treating the full orbital dependence is numerically at least one order of magnitude more complex. It may  however be useful to try to perform some checks, e.g. on the model studied here, to make sure that no other relevant interaction terms depending on more than two orbital indices get generated. The next aspect that should be included and that would add additional precision will be selfenergy corrections and a better treatment of the frequency dependence. The latter will possibly involve form factors also for the frequency dependence. 

Considering the numerical effort, the IOBI or two-orbital approximations with one-frequency parametrization of the frequency dependence should on the same scale (or maybe three or nine times as high) as cRPA or also GW as far as complexity is concerned. One is dealing with effective interactions that depend on one frequency and one momentum and that have a matrix structure in the orbital indices. Therefore, we hope that additional improvements like the inclusion of selfenergy corrections will be feasible. Most importantly, the extension to models with many bands, as may be required to make useful contributions in first-principles theory should become realistic. Turned around, our fRG approach can also be viewed as a coupled-channel generalization of cRPA or GW that adds magnetic and pairing fluctuations to the charge screening physics.

In the cfRG, the effective interactions are obtained by solving the RG differential equations and not by an inversion of a matrix in orbital space as in the cRPA. As the structures that develop in the flow are not extremely sharp, the integration of these coupled differential equations does not pose particular problems. Note that recently, we also reformulated the selfconsistent parquet approximation using the language of a channel-decomposed interaction with form-factor expansions\cite{eckhardt}. The solution of the parquet equations is to a great deal equivalent (or even more complete) than solving the fRG flow equations. By this bridge, we could also reformulate the cfRG as selfconsistent constrained parquet approximation. This would then lead to coupled selfconsistency equations that extend the Bethe-Salpeter-like single-channel equations of the cRPA. Currently, we believe that solving the coupled fRG equations is simpler than solving these coupled selfconsistency equations, but both approaches may find their use.

We acknowledge discussions with S. Andergassen, S. Biermann, J. Ehrlich, C. Hille, J. Hofmann, M. Kinza, L. Markhof, T. Reckling, A. Tagliavini and T. Wehling. The German Research Foundation supported this project, in particular via HO2422/10-1 and HO2422/11-1 and the DFG Research Training Group 1995 ''Quantum Many-Body Methods in Condensed Matter Systems''. Furthermore I am grateful for hospitality by and discussions with the Austrian SFB ViCom, ''Vienna Computational Materials Laboratory''.

\bibliography{iobi}
\end{document}